\documentclass[aps,prb,10pt,floatfix,twocolumn,showkeys,superscriptaddress,longbibliography]{revtex4-1}
\usepackage{graphicx}
\usepackage{xcolor}
\usepackage{amsmath}
\usepackage{bm}
\usepackage{array}
\newcolumntype{H}{>{\setbox0=\hbox\bgroup}c<{\egroup}@{}}
\usepackage[%
  bookmarksopen=true,
  unicode=true,
  hyperfootnotes=true,%
  breaklinks=true,colorlinks=true,linkcolor=blue,urlcolor=blue,citecolor=blue
]
{hyperref}
\usepackage[utf8]{inputenc}
\usepackage{paralist}

\setcitestyle{super}

\input{MyMc}

\newcommand{\Sec}[1]{\section{#1}}

\def\pitaffil{%
  \affiliation{%
    Physikalisches Institut, Center for Quantum Science (CQ) and LISA$^+$,
    Universität Tübingen,
    72076 Tübingen,
    Germany
    }
  }
\def\bsaffil{%
  \affiliation{%
    Institut für Elektrische Messtechnik und Grundlagen der Elektrotechnik,
    Technische Universit\"at Braunschweig,
    38106 Braunschweig,
    Germany
  }
}
\def\zaraffil{%
  \affiliation{%
    Instituto de Nanociencia y Materiales de Aragón (INMA), CSIC-Universidad de Zaragoza,
    50009 Zaragoza,
    Spain
  }
}

\begin{document}

\title{%
  \texorpdfstring{YBa$_2$Cu$_3$O$_7$}{YBa2Cu3O7} nano-constriction Josephson junctions and SQUIDs fabricated by focused helium-ion-beam irradiation
}
\author{Christoph Schmid}
\pitaffil
\author{Christopher Buckreus}
\pitaffil
\author{David Haas}
\pitaffil
\author{Max Pröpper}
\bsaffil
\author{Robin Hutt}
\pitaffil
\author{César Magén}
\zaraffil
\author{Dominik Hanisch}
\bsaffil
\author{Max Karrer}
\pitaffil
\author{Meinhard Schilling}
\bsaffil
\author{Dieter Koelle}
\pitaffil
\author{Reinhold Kleiner}
\pitaffil
\author{Edward Goldobin}
\pitaffil

\date{%
  \today\ File: \textbf{\jobname.\TeX}
}

\begin{abstract}
By focused $30\units{keV}$ He ion beam irradiation, epitaxially grown YBa$_2$Cu$_3$O$_7$ (YBCO) thin films can be driven from the superconducting to the insulating state with increasing irradiation dose.
A properly chosen dose suppresses superconductivity down to 4\,K, while crystallinity is still preserved.
With this approach we create areas of normal-conducting YBCO that can be used to define resistively shunted constriction-type Josephson junctions (cJJs) on the nanometer scale.
We also demonstrate that the fabricated cJJs can be incorporated in direct current superconducting quantum interference devices and can be used as  detector junctions in THz antennas.
\end{abstract}



\maketitle

\Sec{Introduction}
\label{sec:intro}

Constriction Josephson junctions (cJJs) -- sometimes also denoted as superconducting nanowires or Dayem bridges -- embody the paramount necessity when it comes to the down-scaling of Josephson devices below the $100\units{nm}$ limit.
JJs with a barrier usually have a moderate critical current density $j_c$ such that upon down-scaling the junction width, the critical current $I_c$ becomes suppressed by thermal fluctuations
at temperatures $T\approx 4\units{K}$, \ie the Josephson energy $E_J=I_c\Phi_0/2\pi$ becomes comparable to or even less than $k_\mathrm{B}T$; $\Phi_0\approx2.068\units{fWb}$ is the magnetic flux quantum and \(k_\mathrm{B}\) is the Boltzmann constant.
For applications at higher temperatures, \eg, using cuprate superconductors such as YBa$_2$Cu$_3$O$_7$ (YBCO), the problem is even more severe.
Therefore, the best solution to increase $I_c \propto j_c$ is to use barrierless junctions, \ie, cJJs.
Another bonus of cJJs is that they are insensitive to moderate magnetic fields.
Due to these advantages they were already successfully employed in conventional metallic superconductors like Al \cite{vijay2010} and Nb \cite{kennedy2019, uhl2024, Weber25}.
This enabled the fabrication of superconducting quantum intereference nano-devices (nanoSQUIDs) \cite{granata2016, Martinez-Perez17a}
on chip \cite{lam2003}, on lever \cite{wyss2022, Weber25} or on tip \cite{vasyukov2013, anahory2020}, or in photon-pressure devices \cite{bothner2021, rodrigues2021} and qubits \cite{rieger2023}.
One of the most important parameters that define the operation of cJJs and the fabrication requirements is the superconducting coherence length $\xi$.
cJJs with a constriction width $w\approx\xi$ show Josephson-like behavior with sinusoidal current-phase relation (CPR), but for larger $w$ the behavior is inductance-like, with a linear CPR \cite{likharev1979,golubov2004a}.
With that, the fabrication of cJJs in thin films made of  Al ($\xi \approx 100\units{nm}$)
\cite{vanweerdenburg2023, lopez-nunez2025} or Nb ($\xi\approx 10-15\units{nm}$) \cite{pinto2018} is relatively easy or at least technologically feasible.
However, for cuprate high-$T_c$ superconductors with $\xi \approx 1\units{nm}$ the fabrication of cJJs is rather challenging.

Many efforts on the fabrication of YBCO cJJs and nanowires were undertaken during the last decades.
The first significant advancements in fabricating YBCO thin films with nanowires of approximately $200\units{nm}$ in width and length were achieved using electron-beam lithography, metal stencils, and ion beam milling \cite{schneider1993}.
This fabrication technique dominated for decades and enabled the realization of some of the smallest superconducting structures \cite{papari2012, Larsson2000, tafuri2013}, including nanowires as narrow as $50\units{nm}$.
A major breakthrough came from the Chalmers group \cite{nawaz2013}, which demonstrated reproducible YBCO nanowires with a width as small as $40\units{nm}$ and a length of $50\units{nm}$.
This achievement marked a new benchmark and led to a series of further investigations \cite{arpaia2013, arpaia2016, arpaia2017, nawaz2013a, trabaldo2019}.
The same group successfully reduced the nanowire thickness as well, realizing grooved Dayem bridges \cite{trabaldo2020}.
More recently, encapsulating the YBCO thin film with protective layers turned out to be efficient against environmental and processing-induced degradation \cite{ma2025}.
This  provided $68\units{nm}$-wide nanowires with film thickness of only $5\units{nm}$.
Another technically impressive method is the superlattice nanowire pattern transfer (SNAP) technique.
It enabled the fabrication of microscaled nanowire arrays comprising 150 nanowires, each milled down to $15\units{nm}$ wide and $\sim30\units{\mum}$ long having $T_\mathrm{c}\approx20\units{K}$) \cite{xu2008}.

Focused ion beam (FIB) techniques have also proven to be highly effective for YBCO nanostructuring.
Ga-FIB has been used to fabricate nanowires with widths of $55$--$120\units{nm}$ \cite{lam2019, lyatti2018, rouco2018}, as well as to create variable-thickness bridges \cite{wu2015}.
In the context of YBCO grain-boundary JJs \cite{hilgenkamp2002}, miniaturization down to widths of $80\units{nm}$ has also been successfully demonstrated \cite{Nagel11}.
%

He-FIB irradiation represents a promising fabrication technique for modifying YBCO properties on the nanoscale without removing material by milling, but rather by local modification of the electric transport properties of the YBCO films \cite{Cybart17}.
One assumes, that the He ions induce mainly displacements of the loosely bounded chain-oxygens, reducting $T_c$ of YBCO \cite{lang2006, zaluzhnyy2024}.
By irradiating along a line crossing a YBCO thin-film microbridge, typically 2--$4\units{\mu m}$ wide and 30--$50\units{nm}$ thick, using He-FIB one can fabricate a number of elements depending on the irradiation line-dose $D_\ell$ measured in the number of ions per nm of line length (ions/nm) \cite{cybart2015, mueller19,Karrer24,propper2024}.
For moderate line doses $D_\ell$ typically in the range roughly from $400$ up to $1000\units{ions/nm}$, He-FIB irradiation induces a Josephson barrier.
The critical current decreases with $D_\ell$ exponentially and becomes unmeasurable at $4.2\units{K}$ for some $D_\ell>1000\units{ions/nm}$.
In this range one obtains JJs with resistively-shunted junction (RSJ)-like $I$--$V$ characteristics (IVCs) \cite{cybart2015, mueller19}.
For high dose values $D_\ell \gtrsim 1500 \units{ions/nm}$ one obtains highly resistive barriers with fully suppressed Josephson supercurrent at $T=4.2\units{K}$ and above.
Their resistance $R(T)$ increases strongly with decreasing $T$ \cite{mueller19} reaching the \units{M\Omega} range at $4.2\units{K}$.
Scanning transmission electron microscopy (STEM) studies of such  high dose barriers show that the crystal structure is destroyed (amorphized) within irradiated tracks.
We denote such barriers as amorphous insulating barriers (AIBs)\cite{mueller19} and irradiation doses that are high enough to induce amorphization as overcritical doses.
These AIBs have been used for lateral nanopatterning of YBCO thin film devices down to $50\units{nm}$ linewidth \cite{cho2018}.
Accordingly, AIBs can be used for defining the boundaries of electronic circuits and (SQUID) holes on the nanoscale.
Various devices, with AIBs and JJs that were all written during one He-FIB run, have been demonstrated, \eg, SQUIDs \cite{mueller19, cho2018, cho2018a, li2019} or Josephson diodes \cite{Schmid:2025:He-FIB:YBCO-JJD}.

An obvious approach to create ultra-narrow YBCO cJJs with widths down to only a few nm could rely on the use of AIBs.
However, the lateral damage produced outside the He-FIB-induced amporphous tracks is significant on a length scale $\sim50\units{nm}$ on both sides, see Appendix \ref{app:aiw} for details.
In this paper we present an alternative strategy to define ultra-narrow YBCO cJJs, approaching widths down to the FIB spot size of a few nm.
Instead of narrow high-dose lines, we use He-FIB-induced resistive areas.
The idea is to use doses that are equivalent to the moderate doses used for creating Josephson barriers, in the sense that amorphization is avoided.
However, now the irradiated regions with suppressed superconductivity have an extension along current direction that is too large to allow for Cooper pair tunneling.
This approach shall minimize lateral damage and hence bring the minimal width of a constriction roughly down to the He-FIB spot size.

\Sec{Thin film Fabrication and prepatterning}
\label{sec:fabrication}

$c$-axis oriented YBCO thin films with thickness $t = 30 \units{nm}$ are epitaxially grown on a $10 \times 10 \units{mm^2}$ \LSAT (LSAT) substrate by pulsed laser deposition \cite{schilling2013}.
An ex-situ Ag ($30\units{nm}$) plus Au ($20\units{nm}$) capping layer is deposited and used for contact pads later on.
Using optical lithography and Ar ion milling , the YBCO-Ag-Au trilayer is prepatterned into 144 microbridges, all $100\units{\mum}$ long and $2\units{\mum}$ wide connected on one side to a common ground and on the other to individual voltage and current pads.
Then, using wet-etching process by means of TechniEtch ACI2 from MicroChemicals, the capping layer is removed in the areas to be irradiated by He-FIB and also between the contact pads to allow for 4-point transport measurements. In total, four of these chips were investigated.

\begin{figure}
  \includegraphics[width=0.96\columnwidth]{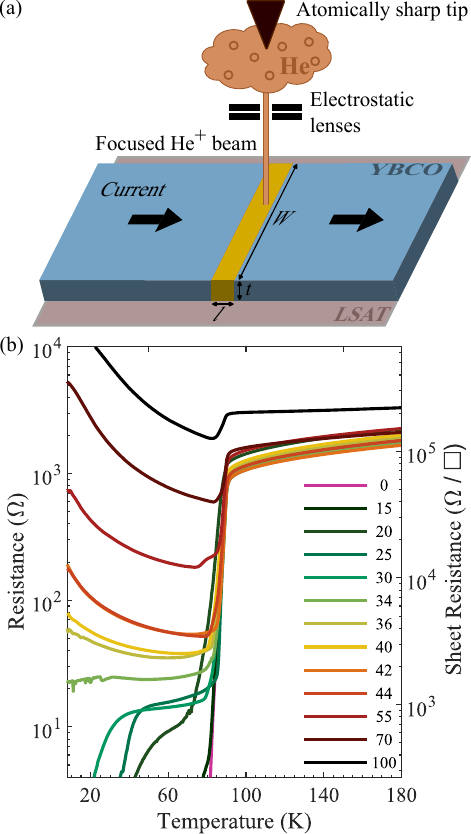}
  \caption{%
    (a) Sketch of an YBCO microbridge with He-FIB-irradiated area of length $\ell$ and width $W$.
    The YBCO film is $t = 30 \units{nm}$ thick.
    (b) $R(T)$ curves of microbridges that contain areas ($\ell=30\,$nm, $W=2\units{\mum}$) irradiated with different areal doses $D_\mathrm{a}$ (color indicates $D_\mathrm{a}$ in $\units{ions/nm^2}$).
    $R$ is measured at constant bias current $I = 1 \units{\mu A}$.
    The right vertical axis shows \(R\,W / l\), which below $T_\mathrm{c,0}$ corresponds to the sheet resistance \(R_\square\) of the irradiated region.
  }
  \label{fig:rt}
\end{figure}

\Sec{Resistive areas written by He-FIB}
\label{sec:resistors}

For the investigation of the resistive areas, we irradiated a set of microbridges, each with a rectangular area across it, as shown in Fig.~\ref{fig:rt}(a).
For irradiation we used  a He-FIB pixel spacing (pitch) \(p = 1 \units{nm}\).
We assume that the He-FIB has a Gaussian profile with a standard deviation $\sigma \approx 3\units{nm}$, extracted from the STEM data in Ref.~\onlinecite{mueller19}.
With that, the value for the pitch is a few times smaller than the estimated He-FIB spot-size $2\sigma$.
Each microbridge was irradiated with a different areal dose $D_\mathrm{a}$ ranging from 0 to $100\units{ions/nm^2}$.
The measured $R(T)$ curves of the microbridges are shown in Fig.~\ref{fig:rt}(b).
These curves demonstrate that He ion irradiation results in a superconductor-to-insulator transition as the dose increases, which confirms previous works \cite{lang2006, sarkar2023, zaluzhnyy2024}.
For low dose values up to 30\,ions/nm$^2$, $T_c$  decreases with increasing dose, shifting towards $8\units{K}$ -- our lowest measurement temperature for $R(T)$ measurements.
At $D_\mathrm{a}\approx 34\units{ions/nm^2}$ we do not observe a superconducting transition anymore and below  $T_{c,0}\approx 89\units{K}$ --- the critical temperature of the unirradiated part of the YBCO microbridge --- the resistance stays almost constant at $R\approx 22\,\Omega$.
Assuming that below $T_{c,0}$ the resistance is entirely due to the irradiated region, we can estimate its sheet resistance as $R_\square= RW/\ell\approx 1.5\units{k\Omega/\square}$, where the symbol $\square$ denotes a geometrical square area of thin film.
With further increase of the dose, the resistance below  $T_{c,0}$ increases and $R(T)$ becomes semiconductor-like reaching very high values at low $T$.
%
Note that $R(D_a)$ curves at fixed $T<T_\mathrm{c,0}$ show an exponential increase with $D_a$.
This is probably compatible with Anderson localization and variable range hopping \cite{Lesueur:1993:YBCO:SIT(He-ions)}.
From our point of view, the most attractive dose for fabricaton of the resistors is around 35--$40\units{ions/nm^2}$.
Here the values of the resistance are lowest and almost do not depend on $T$ up to $T_\mathrm{c,0}$.
These YBCO resistors can also serve as circuit components, allowing straightforward integration into He-FIB–fabricated devices.

\Sec{Constriction Josephson Junctions}
\label{sec:cJJs}

\newcommand{\InsetGeom}{top-left\xspace}
\newcommand{\InsetDeff}{bottom-right\xspace}

\begin{figure}
  \includegraphics[width=0.95\columnwidth]{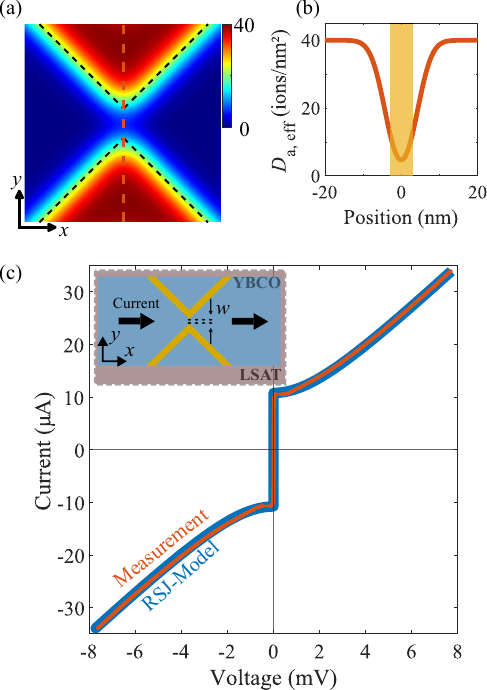}
  \caption{%
YBCO cJJ:
    (a)~Color plot $D_\mathrm{a,eff}(x,y)$ calculated for a constriction with $w=6\units{nm}$ and for a He-FIB with $\sigma=3\units{nm}$ in the $40 \times 40\units{nm^2}$ region around the constriction.
    The dashed lines indicate the nominal edges of the irradiated areas (triangles) with the design distance \(w = 6 \units{nm}\) between the tips.
    The  $D_\mathrm{a,eff}(0,y)$ profile indicated by the orange dashed line is shown in (b).
    The yellow area indicates the cJJ width $w$.
    (c)~The orange IVC describes a cJJ irradiated with $D_\mathrm{a} = 40 \units{ions/nm^2}$ and $w = 6 \units{nm}$, measured at $T = 4.2 \units{K}$.
    The blue IVC illustrates a RSJ-fit with a critical current \(I_\mathrm{c} = 10.8 \units{\mu A} \) and a junction resistance \(R_\mathrm{n} = 240 \units{\Omega}\).
    The inset is a sketch of the bridge geometry, where the yellow regions indicate the He-FIB-irradiated resistive regions (100 nm wide) that define the banks of a cJJ with  the nominal constriction width $w$.
  }
  \label{fig:cjj}
\end{figure}

To define cJJs with a constriction width well below $100\units{nm}$ we use He-FIB-induced YBCO resistors described in Sec.~\ref{sec:resistors}.
We create two triangular resistive areas with their tips almost touching each other, see the inset of Fig.~\ref{fig:cjj}(c).
The distance between the tips defines the nominal constriction width $w$.
For constrictions with $w \lesssim 30\units{nm}$, the finite He-FIB spot size should be taken into account --- the edges of the resistive areas are smeared on a scale $\sim\sigma$ of the He-FIB with a Gaussian cross-section profile.
An effective areal dose $D_\mathrm{a,eff}(x,y)$ in each point can then be calculated as the convolution of the Gaussian beam profile with the pixel map defining the irradiation pattern.
Figure \ref{fig:cjj}(a) shows an example of a $D_\mathrm{a,eff}(x,y)$ profile, with Fig.~\ref{fig:cjj}(b) presenting the corresponding cross-sectional cut through the constriction.
Obviously, at the tips of the resistive areas the effective dose is somewhat lower than inside the resistive triangles.
However, the dose inside the constriction is also non-vanishing as nominally assumed.
Thus, the critical current density distribution $j_c(0,y)$ across the constriction may significantly deviate from a step-like profile, and we expect smearing of this profile on a length scale of several $\sigma$.
Thus, the effective constriction width (defined as the region along the $(0,y)$) line where $T_c(y)>T$ can vary as a function of $w$ and $D_a$ and $T$ and can even be made smaller than $2\sigma$.
Relying on the results presented in section \ref{sec:resistors}, we fabricated a set of cJJs using a dose of $D_\mathrm{a} = 40 \units{ions/nm^2}$ to create the resistive banks of cJJs.
This dose is undercritical in terms of amorphization and provides an approximately constant resistivity over the whole temperature range from $5\units{K}$ to $T_\mathrm{c,0}$.

\begin{figure}
  \centering
  \includegraphics{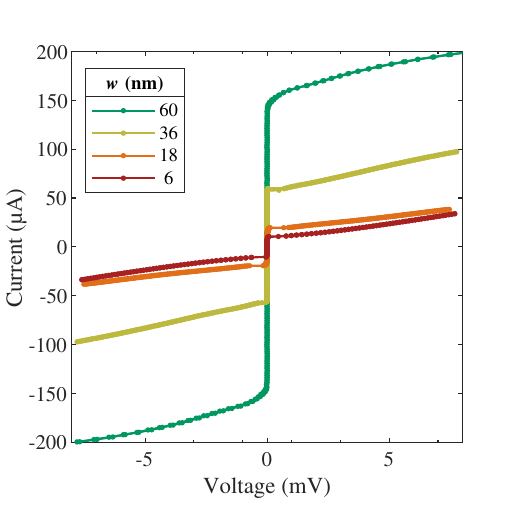}
  \caption{%
    IVCs (at $T = 4.2\,$K) of YBCO cJJs with different constriction width $w$ all irradiated with $D_\mathrm{a} = 40 \units{ions/nm^2}$.
    %
  } 
  \label{fig:IVCs(w)}
\end{figure}

Using the geometry depicted in the inset of Fig.~\ref{fig:cjj}(c), we fabricated a series of cJJs with nominal constriction width $w$ from 0 to $66\units{nm}$ and characterized their electric transport properties at $T=4.2\units{K}$.
Our best result in terms of downscaling is a cJJ with \(w = 6 \units{nm}\) showing an RSJ-like IVC, see Fig.~\ref{fig:cjj}(c).
IVCs of several other cJJs with different $w$ are presented in Fig.~\ref{fig:IVCs(w)}.
%
%
For increasing $w$ the IVCs show a significant excess current $I_\mathrm{ex}$ and a linear, rather than RSJ-like hyperbolic branch above $I_\mathrm{c}$, typical for superconductor-constriction-superconductor systems.
To determine the junction resistance $R_\mathrm{n}$,  the slope of the IVC in the linear regime well above $I_\mathrm{c}$ is determined by fitting with $I=V/R_\mathrm{n}+I_\mathrm{ex}$ (not shown in the graph)
and using $R_\mathrm{n}$ and $I_\mathrm{ex}$ as fitting parameters.
At $w\gtrsim 48\units{nm}$, the shape of the IVCs becomes flux-flow-like with a substantial rounding near $I_\mathrm{c}$.
The value of $I_\mathrm{c}$ was determined with a voltage criterion of $8 \units{\mu V}$.

\begin{figure}
  \centering
  \includegraphics{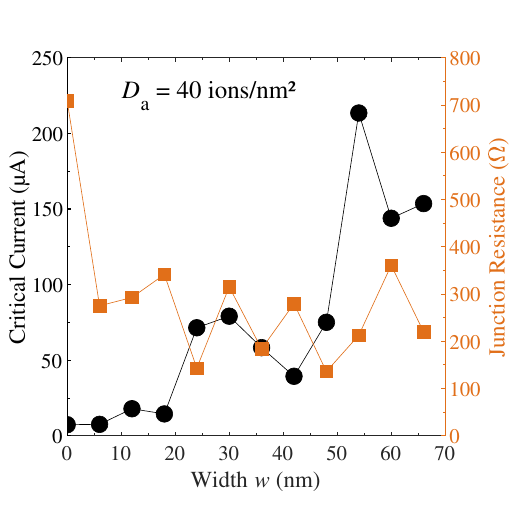}
  \caption{%
    Critical current $I_c(w)$ and junction resistance $R_\mathrm{n}(w)$ for YBCO cJJs irradiated with $D_\mathrm{a} = 40 \units{ions/nm^2}$, extracted from IVCs measured at \(T = 4.2 \units{K}\).
  }
  \label{fig:Ic&R(w)}
\end{figure}

In Fig.~\ref{fig:Ic&R(w)} we plot $I_c(w)$ and $R_\mathrm{n}(w)$ to summarize our findings.
We observe a substantial spread in the cJJ parameters, similar to (or even larger than in) He-FIB barrier Josephson junctions \cite{mueller19,propper2025}.
While  $I_c$ shows an overall increase with increasing $w$, the resistance $R_\mathrm{n}$ stays more or less constant on the scale of the observed scattering.
The reason for that is unknwon and under further investigation.

To test the high-frequency properties of the cJJs, in view of possible applications for YBCO voltage standards, we fabricated several cJJs in lithographically structured antennas.
Under excitation with a far-infrared laser, we successfully observed Shapiro steps in the IVC. See Appendix~\ref{app:thz} for details.
We investigated IVCs at different $T$.
The investigated cJJs demonstrate reliable operation over a wide temperature range and closely follow the theoretical model proposed by Bardeen \cite{bardeen1962}.
For additional details, see Appendix~\ref{app:ict}.

\section{dc SQUID with two constriction junctions}
\label{sec:SQUIDs}

\begin{figure*}
  \includegraphics[width=0.8\textwidth]{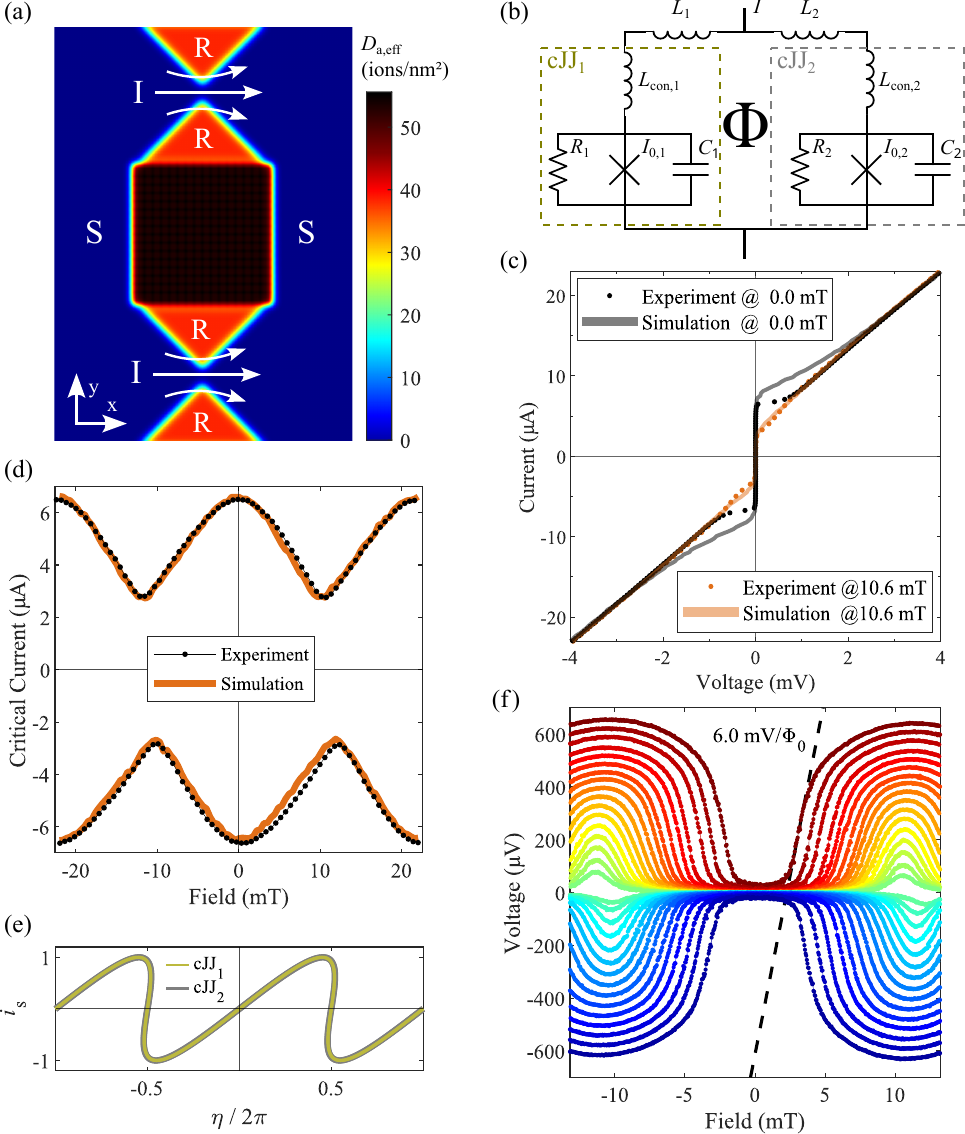}
  \caption{%
YBCO dc SQUID with cJJs:
    (a) Calculated effective dose $D_\mathrm{a,eff}(x,y)$, assuming a Gaussian distribution of the He-FIB intensity with $\sigma=3 \units{nm}$.
    (b) Circuit diagram of the dc SQUID used for simulations.
    (c) Experimental (dots) and simulated (lines) IVCs of the nanoSQUID at zero field (black) and at \(B=10.6 \units{mT}\) (orange).
    (d) Experimental (black) and simulated (orange) field modulation of positive and negative critical current.
    (e) Normalized current phase relations of the weak links, as extracted from simulations.
    (f) Voltage \vs applied magnetic field dependence at different bias currents through the nanoSQUID from \(I = -6.33 \units{\mu A}\) (dark blue) to \(I = 6.35 \units{\mu A}\) (dark red)).
    All measurements were performed at \(T\approx 4.2\units{K}\).
  }
  \label{fig:squid}
\end{figure*}

Using the technique described above, we produced direct current (dc) nanoSQUIDs.
Within a \(2 \units{\mu m}\) wide microbridge, we centered a $100 \times 100 \units{nm^2}$ SQUID hole irradiated by He-FIB with a dose of \(D_\mathrm{a} = 56 \units{ions/nm^2}\), to make sure that superconductivity is well suppressed.
We used a pitch value of \(p = 6 \units{nm}\), which results in $D_\mathrm{eff}(x,y)$ shown in Fig.~\ref{fig:squid}(a).
On two sides of the hole we created two cJJs as described in the previous section with \(D_\mathrm{a} = 40 \units{ions/nm^2}\) (with finer \(p = 1\units{nm}\)) and \(w = 12 \units{nm}\).
%
%
The cJJs are placed at a distance of \(\approx 50\units{nm}\) from the edge of the SQUID-hole to reduce the impact of mechanical stress (if any) on the constrictions.
%
Figure \ref{fig:squid}(b) shows the circuit diagram of the SQUID, as outlined in more detail below.
The IVC of the nanoSQUID, see Fig.~\ref{fig:squid}(c), exhibits a maximum critical current of about \(6.5 \units{\mu A}\) and \(R_\mathrm{n} \approx 210 \units{\Omega}\),
at zero applied magnetic field perpendicular to the sample plane.
The first minimum of $I_c(B)$  is obtained at $B=10.6 \units{mT}$, where $I_c=2.8 \units{\mu A}$.
The IVC at this field is also shown in Fig.~\ref{fig:squid}(c).
The values of $I_c$ are determined by using a 50 \units{\mu V} voltage criterion.
The $I_c(B)$ modulation pattern, see Fig.~\ref{fig:squid}(d), reveals an almost symmetric modulation with a modulation depth of \(\Delta I_c/I_c \approx 58 \% \).

From the period $\Delta B\approx 22.4 \units{mT}$ of the $I_c(B)$-modulation the effective area of the SQUID is estimated as
\(A_\mathrm{eff} \approx 9\cdot10^4 \units{nm}^2 \), which is 6 times larger than the physical hole area $A_\mathrm{hole}=1.5\cdot10^4\units{nm}^2$, including the triangles forming the cJJ banks.
This kind of flux focusing is typical for planar thin film structures.
Numerical simulation of effective area
\footnote{Effective area $A_\mathrm{eff}$ is calculated as fluxoid $\tilde{\Phi}$ induced around the SQUID hole (output of the simulation) divided by the applied field $B$ (input parameter), \ie, $A_\mathrm{eff}=\tilde{\Phi}/B$.}
using 3D-MLSI \cite{khapaev2001} gives a 14 times larger area than $A_\mathrm{hole}$.
The origin of the discrepancy between measured and calculated flux focusing factor is currently not known.
We have also measured $V(B)$ curves taken at different bias currents, see Fig.~\ref{fig:squid}(f).
The best transfer function of \(6.0 \units{mV / \Phi_0}\) was obtained at \(I=6.35 \units{\mu A}\).

\newcommand{\betaLcon}[1]{%
\ensuremath{\beta_{\mathrm{L},#1}^\mathrm{con}}\xspace%
}
\newcommand{\lcon}[1]{%
\ensuremath{l_{#1}^\mathrm{con}}\xspace%
}

To compare our data with theoretical expectations and to exctract some unknown parameters we start from the circuit diagram in Fig.~\ref{fig:squid}(b) and in essence solve the corresponding standard Langevin equations for the phase dynamics \cite{tesche1977}.
In the model each of the two SQUID arms contains a JJ with the noise free critical current \(I_\mathrm{0,1}\) (or \(I_\mathrm{0,2}\)), which is shunted by a resistor \(R_\mathrm{1}\) (or \(R_\mathrm{2}\)), which also generates a white noise according to the ambient temperature $T$ and a capacitor \(C_\mathrm{1}\) (or \(C_\mathrm{2}\)).
The inductances of the SQUID arms are \(L_\mathrm{1}\) and \(L_\mathrm{2}\), respectively.
In the standard model, the current-phase-relation of both JJs is assumed to be sinusoidal.
In our case we expect that the CPR is skewed.
Unfortunately, there is no unique expression describing the CPR’s of all weak links microscopically.
However, Likharev \cite{likharev1979} proposed to model the constriction by a serial connection of an inductor \(L_\mathrm{con}\) and an ideal JJ with sinusoidal CPR.
We use this ansatz for our simulations. i.e. the SQUID dynamics can be described by the sinusoidal CPRs, but with an effective inductance of each SQUID arm $L_\mathrm{eff,1}=L_1+L_\mathrm{con,1}$ and $L_\mathrm{eff,2}=L_2+L_\mathrm{con,2}$.
We then solve the dynamic Langevin equations for the Josephson phase differences $\phi_1$ and $\phi_2$ and calculate the dc voltages using the Josephson relation.
We finally obtain the SQUID critical current by applying a proper voltage criterion.
The simulations are performed in normalized units.
We introduce the average JJ critical current \(I_0 = (I_{0,1}+I_{0,2})/2 \) and normalize other quantities using $I_0$.
The parameters entering the model are the critical currents  \(I_{0,1}\) and  \(I_{0,2}\) \ie, \(i_{0,1}\) and  \(i_{0,2}\) in normalized units, the normalized SQUID inductance \(\beta_L = 2 I_0 (L_1 + L_2 ) / \Phi_0 \), the noise parameter \(\Gamma=  2\pi k_\mathrm{B} T / I_0\Phi_0 \).
Further, the normalized constriction inductances are \(\betaLcon{1}= 2 I_0 L_\mathrm{con,1} / \Phi_0 \) and \(\betaLcon{2}=2 I_0 L_\mathrm{con,2} / \Phi_0 \) and the Stewart-McCumber parameter \(\beta_C = 2 \pi I_0 R^2 C / \Phi_0 \), \(R\) is the average junction resistance and \(C\) is the average junction capacitance.
As a result of fitting $I_c(B)$ we find $I_0\approx4.2 \units{\mu A}$, resulting in \(\Gamma=0.043\).
The SQUID inductance $L_1+L_2$ is estimated using 3D-MLSI \cite{khapaev2001}, resulting in a value of $23.6\units{pH}$ and leading to \(\beta_\mathrm{L} \approx 0.1\).

This value of $\beta_\mathrm{L}$ is much smaller than the effective dimensionless SQUID inductance $\beta\mathrm{_L^{eff}}=\beta_\mathrm{L}+\betaLcon{1}+\betaLcon{2}\approx0.9$ required for the best fit to the data.
Thus, we conclude that the SQUID inductance is dominated by the constriction inductances.
3D-MLSI, solving the London equations, assumes a spatially constant Cooper pair density $n_\mathrm{s}$ and thus a spatially constant London penetration depth $\lambda_\mathrm{L}$.
In the constriction regions $n_\mathrm{s}$ is likely to be reduced, and in addition microscopic processes like Andreev reflections may contribute to the distortion of the CPRs, leading to large values of $L_\mathrm{con,1}$ and $L_\mathrm{con,2}$.
Further, for the best fit we used \(i_{0,1} = I_{0,1}/I_0 = 1.1\) and \(i_{0,2} = I_{0,2} / I_0 = 0.9.\)
The normalized constriction inductances were $\betaLcon{1}=0.364$ and $\betaLcon{2}=0.454$.
For simplicity we assumed that the junction resistances are inversely proportional to \(i_{0,1}\) and \(i_{0,2}\), \ie, we kept the product of the junction resistances and critical currents constant.
We further assumed equal junction capacitances and used a Stewart-McCumber parameter \(\beta_C =0.2 \).

Figure \ref{fig:squid}(c) shows (solid lines) the resulting simulated IVCs for both maximum and minimum critical current.
The experimental curve for minimum critical current is reproduced well, while there are deviations for the IVC at maximum critical current.
An increase of \(\beta_C\) to values around 1 would have caused a better agreement with this curve, however, would lead to a strong $LC$ resonance visible as a bump in the IVC for $B=10.6\units{mT}$ crossing the zero-field IVC in some voltage range, which is not observed.
Thus, we rule out large values of \(\beta_C\) by means of pure fitting and also note that such a scenario would not match our weak-link geometry, where we expect at best very small stray capacitances.
We assume that the significant deviations of the zero-field IVC arise from Joule heating or other nonequilibrium effects causing a suppression of the superfluid density in the weak link region.
Finally, a comparison of the high-voltage regime of the experimental and simulated IVCs yields a characteristic voltage \(I_0 R\) of about $1.5\units{mV}$, resulting in a dimensionless voltage criterion of 0.033, which was used for simulation of $I_c(B)$.
Fig.~\ref{fig:squid}(d) shows the simulated $I_c(B)$ patterns by the orange lines, both for positive and negative polarities.
For the comparison the experimental data needed to be shifted slightly by 0.02 \(\Phi_0\) on the flux axis to account for residual magnetic fields in the setup.
The agreement between the experimental curves and the simulated ones is good.
Note that the data for negative currents are reproduced less well.
This is, because the experimental curves are not exactly point symmetric while the simulated ones are.
The fits have been made for postitive currents.
Overall, the essential parameters for the \(I_c(B)\) patterns (junction asymmetries and constriction inductances) leave very little degrees of freedom, only a few percent.
In Fig.~\ref{fig:squid}(e), we show the resulting skewed CPR, resulting from Likharev’s ansatz, leading to the implicit expression.
\begin{equation}
  I_k = I_{0,k} \sin \braces{ \eta_k-\frac{2\pi}{\Phi_0} L_{\mathrm{con}, k} I_k} ,\text{ where }k = 1,2
  , \label{Eq:CPR.imp}
\end{equation}
where \(I_k\) are the currents through the cJJs. The phases \(\eta_k\) are the sum of the phase drops across the constriction inductances and the ideal junctions, and the argument of the sine function yields the phase across the ideal JJ. In dimensionless units
\begin{equation}
  i_k = \sin \braces{\eta_k - \pi \betaLcon{k} i_{0,k} i_k }
      = \sin \braces{\eta_k - \pi \lcon{k} i_k }
  , \label{Eq:CPR.imp.norm}
\end{equation}
where $i_k=I_k/I_{0,k}$ is the bias current normalized to the critical current of \emph{each} JJ and $\lcon{k} = 2L_{\mathrm{con}, k} I_{0,k}/\Phi_0$ are normilized inductances.

If, in Eq.~\eqref{Eq:CPR.imp.norm}, the parameter \(\pi \lcon{k}\equiv\pi \betaLcon{k} i_{0, k} > 1 \), then the CPR becomes multivalued.
In our case we have \(\pi\lcon{1}=\pi\cdot0.364\cdot1.1\approx1.26\) and \(\pi\lcon{2}\approx\pi\cdot0.454\cdot0.9\approx1.28\) for both JJs.
The corresponding normalized CPRs are plotted in Fig.~\ref{fig:squid}(e), which are nearly identical.

\Sec{Conclusions}
\label{sec:conclusions}

Using resistive areas created in a YBCO thin film by He-FIB irradiation, we fabricated cJJs with the nominal constriction size as small as $w=6\units{nm}$ and demonstrated its operation.
For smallest $w$ we observed an RSJ-like IVC with critical current $\sim10\units{\mu A}$ and characteristic voltage $V_c\approx2.4\units{mV}$ and no excess current at 4.2\,K.
For larger $w$ excess current appears and dominates for $w>20\units{nm}$.
Flux-flow like IVCs are observed for $w\approx 50\units{nm}$ and above.
$I_c(T)$ dependences demonstrate that cJJs can be used at temperatures up to $50$--$70\units{K}$.
However, for this the constriction cannot be too small and the IVCs are not RSJ-like.

Further, we demonstrated the realization of a dc nanoSQUID, where both the two cJJs and the SQUID hole are written by He-FIB irradiation within a $2\units{\mum}$ wide YBCO microbridge.
The SQUID shows a $I_\mathrm{c}(B)$ modulation with the depth 58\%, which can be explained by an additional constriction inductance.
The $V(B)$ curves demonstrate a transfer function up to $6\units{mV}/\Phi_0$.
The size of the SQUID is defined by the microbridge width and can be further reduced.

\acknowledgments

This work was funded by the Deutsche Forschungsgemeischaft (DFG) via project No.~GO-1106/6 and GO-1106/7
and by the European Commission under H2020 FET Open grant “FIBsuperProbes” (Grant No. 892427). We also gratefully
acknowledge support by the COST actions NANOCOHYBRI (CA16218), FIT4NANO (CA19140) and SUPERQUMAP (CA21144).
We thank M.~Turad and R.~L\"offler for invaluable help with ``Orion Nanofab''.
The authors acknowledge the use of instrumentation as well as the technical advice provided by the National Facility ELECMI ICTS, at the node Laboratorio de Microscopias Avanzadas (LMA) of the Universidad de Zaragoza.

\appendix

\Sec{cJJs created by AIBs}
\label{app:aiw}

\begin{figure}
  \includegraphics[width=0.96\columnwidth]{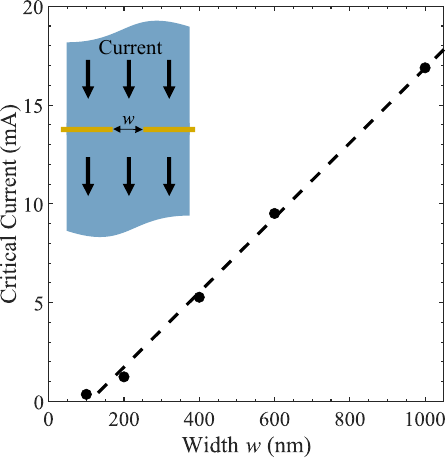}\\
  \caption{%
    Experimentally measured $I_c(w)$ of  cJJs fabricated with He-FIB-induced AIBs with \(D=3000 \units{ions/nm} \) (dots). The dashed line shows a  fit by the linear function $I_c(w)=J_c\cdot (w-w_0)$, which gives $J_c=20\units{\mu A/nm}$ and $w_0=108\units{nm}$.
    The inset depicts the cJJs geometry formed by AIBs (horizontal yellow lines).
  }
  \label{fig:AWR-cJJ:Ic(w)}
\end{figure}

\begin{figure}
  \centering
  \includegraphics{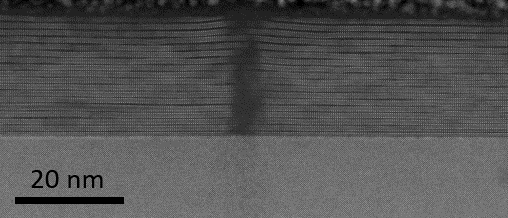}
  \caption{%
    Cross-sectional HAADF-STEM image of a YBCO lamella cut across (perpendicular to) the single line irradiated by He-FIB with a dose $D=2000\units{ions/nm}$.
    The bottom layer is  LSAT, the middle layer is YBCO and the top layer is a Pt-C deposit grown by focused electron beam induced deposition to protect the YBCO film during TEM lamella preparation.
    In the YBCO layer one sees atomic planes, while the dark vertical column in the center corresponds to the amorphized YBCO.
  }
  \label{fig:STEM}
\end{figure}

\begin{figure*}
  \includegraphics{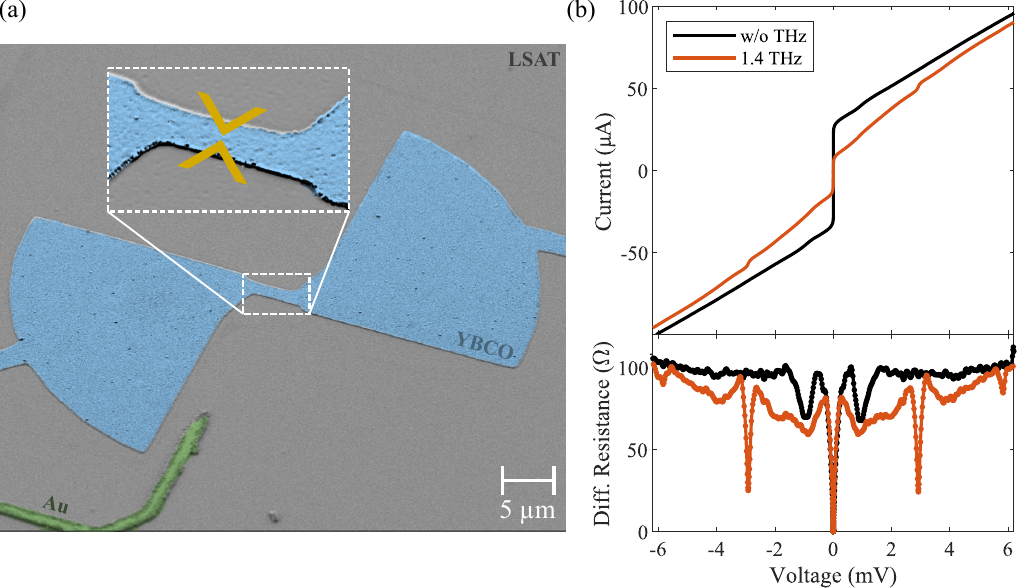}
  \caption{%
    (a) False-colored scanning electron microscopy image of the bow-tie antenna.
    The antenna has a $2\units{\mu m}$ wide microbridge at the center, which is depicted in the inset.
    The yellow triangles indicate the He-FIB irradiation pattern which is used to create the $8\units{nm}$ wide cJJ.
    The thin Au covered trace entering from the bottom left edge of the picture is used as a sacrificial structure for the He-FIB.
    (b) IVC and \(R_\mathrm{d}(V)\) of the cJJ embedded into the antenna structure at \(T = 20 \units{K}\).
    Black curves are measured without and orange curves with $1.4\units{THz}$ excitation.
  }
  \label{fig:antenna}
\end{figure*}

Our first effort to make cJJs is illustrated in the inset of Fig.~\ref{fig:AWR-cJJ:Ic(w)}.
An AIB with \(D=3000 \units{ions/nm} \) was drawn by He-FIB almost across the whole microbridge, except for a constriction of width $w=1000,\,500\,\ldots,50\units{nm}$.
Then, the critical current of each cJJ was measured and plotted as a function of $w$, see Fig.~\ref{fig:AWR-cJJ:Ic(w)}.
$I_c(w)$, as expected, decreases linearly with $w$.
However, the cJJs with $w<100\units{nm}$ (no data points shown) not only showed no critical current but had very high resistance, which is an indication of transport interruption through the constriction.
To make sure that this is not caused by accidentally broken cJJs, we also made a linear fit of $I_c(w)$ data using the function
\begin{equation}
  I_c(w) = J_c(w-w_0)
  , \label{Eq:Ic(w).fit}
\end{equation}
see Fig.~\ref{fig:AWR-cJJ:Ic(w)}, where \(J_\mathrm{c}\) is the critical current per junction width.
The fit shows that the linear dependence does not arrive at the origin, but is shifted by $w_0\approx100\units{nm}$.
Thus, for all measured cJJs the effective width $w-w_0$ is systematically smaller then the design width $w$.

High-angle annular dark field (HAADF) STEM analysis of AIBs, illustrated in Fig.~\ref{fig:STEM}, was performed in a Thermo Fisher Scientific probe-corrected Titan 60-300, equipped with a CETCOR aberration corrector for the condenser systems, with a STEM probe size smaller than 1\units{\AA}.
The cross sectional lamella specimen was prepared in a Thermo Fisher Scientific Helios 650 Dual Beam.
The atomic resolution image evidences the bending STEM analysis of AIBs, see Fig.~\ref{fig:STEM}, showing a bending of YBCO atomic planes near amorphized region towards the surface of the sample.
Obviously, in the amorphized region the YBCO atoms are not optimally packed.
This, probably, leads to a mechanical pressure onto the neighboring still crystalline regions of YBCO, which results in the up-bending.
Visually, the bending takes place in an approximately $50\units{nm}$ wide lateral vicinity on both sides of the AIB, which suggests that superconductivity is suppressed in this region.
Sometimes one also observes cracks (not shown).
Thus, AIBs are not suitable for the fabrication of nano-cJJs or any other nanostructures with lateral sizes  $\lesssim100\units{nm}$.

\Sec{cJJ under THz-Irradiation}
\label{app:thz}

The cJJs presented in section \ref{sec:cJJs} have rather high characteristic voltages $V_c > 2.5 \units{mV}$, making them attractive for THz applications such as voltage standards or detectors \cite{behr2012, rufenacht2018a, propper2024}.
Here we demonstrate the appearance of Shapiro steps under irradiation in the THz range.

To perform this experiment, we have used a YBCO film  with a microstructured bow-tie antenna, see Fig.~\ref{fig:antenna}(a).
Then, using the procedure described above, we have patterned a cJJ with $w = 8\units{nm}$ at the feed point of the antenna.
We used \(D_\mathrm{c} = 40 \units{ions/nm^2}\) and \(p = 1 \units{nm}\) to draw the constriction banks.
At $T=20\units{K}$ we obtained \(I_\mathrm{c} = 28 \units{\mu A}\) and normal resistance of \(R_\mathrm{n} \approx 100\units{\Ohm}\), see Fig.~\ref{fig:antenna}(b).

The cJJ is investigated then as a THz detector, which operates in the frequency range of the far-infrared laser FIRL-100 from Edinburgh Instruments.
In our particular case \cite{tollkuhn2022, ritter2022} the available laser emits simulataneosly at the frequencies of $1$, $1.3$ and $1.4\units{THz}$.
By adjusting the position of the cJJ relative to the laser beam, one can choose the dominating frequency, which couples into the THz antenna.
Under laser irradiation the IVC of the cJJ exhibits a smaller critical current and reveals a small Shapiro step at $2.9\units{mV}$, see Fig.~\ref{fig:antenna}(b).
This exactly corresponds to the $1.4 \units{THz}$ mode of the laser.
Calculating the differential resistance $R_d(V)$ from the IVC and comparing the peaks of the $R_d(V)$ obtained with and without irradiation allows us to observe even the second Shapiro step, which is hardly visible in the IVC.
The appearance of Shapiro steps indicates that high-frequency radiation can be effectively coupled to cJJs and that these junctions display real Josephson (nonlinear) behavior, rather than merely acting as linear inductors.

\begin{figure}
  \includegraphics{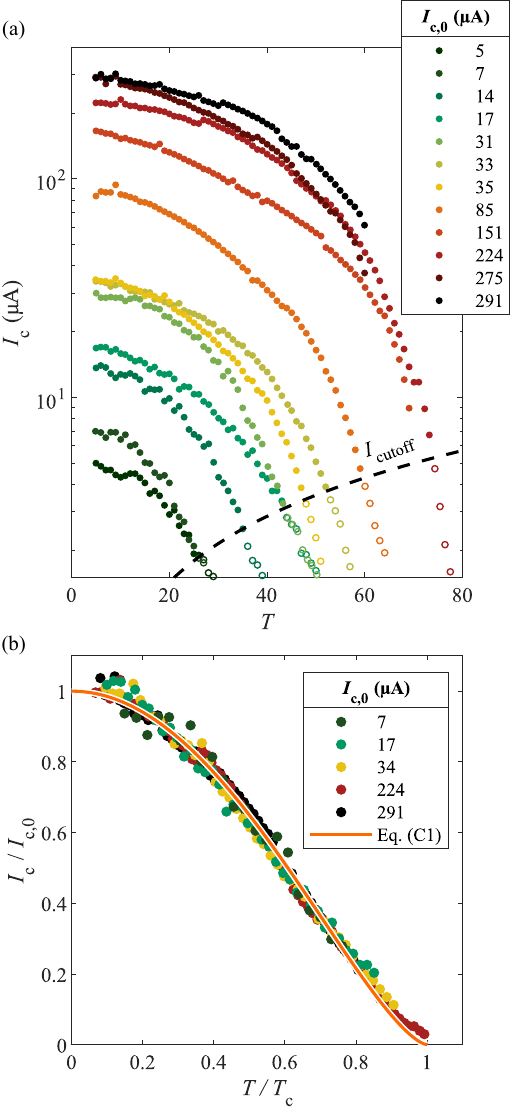}
  \caption{%
    (a) Experimentally measured $I_c(T)$ dependencies of the devices from Tab.~\ref{tab:ict}.
    The cut-off criterion $I_\mathrm{cutoff}(T)$ (dashed line) defines the fitting range for Eq.~\eqref{eq:fiteq}.
    The corresponding best-fit parameters for the considered data points (dots) are summarized in Tab.~\ref{tab:ict}, while the omitted data are marked by circles.
    The legend shows the depairing currents $I_\mathrm{c,0}$.
    (b)
    A subset of the data from (a), showing the normalized critical current $I_\mathrm{c} / I_\mathrm{c,0}$ versus the normalized temperature $T/T_\mathrm{c}$.
    The normalization factors $I_\mathrm{c,0}$ and $T_\mathrm{c}$ corresponds to the best-fit values and are listed in Tab.~\ref{tab:ict}.
    Eq.~\eqref{eq:fiteq} is represented by the continuous orange line.
  }
  \label{fig:ict}
\end{figure}

\begin{table}
    \begin{tabular}{l|l|l|l|l|l|l}
      $I_{\mathrm{c,0}}$ & $T_\mathrm{c}$ & $j_{\mathrm{c,0}} $  & $A$ & $w$  & $D_\mathrm{a}$  & cJJ  \\
      ($\mu$A) & (\units{\K}) & (\units{MA/cm^2})  & (\units{nm^2}) &  (nm) & (\units{ions/nm^2})  &   \\
      \hline
      5 & 37 & 3 & 180 & 6 & 47 & A2 \\
      7 & 33 & 2 & 480 & 16 & 40 & A12 \\
      14 & 42 & 8 & 180 & 6 & 47 & A1 \\
      17 & 51 & 9 & 180 & 6 & 40 & A4 \\
      31 & 48 & 17 & 180 & 6 & 40 & A3 \\
      33 & 58 & 14 & 240 & 8 & 40 & A5 \\
      34 & 52 & 10 & 360 & 12 & 47 & A8 \\
      84 & 59 & 35 & 240 & 8 & 40 & A6 \\
      150 & 72 & 31 & 480 & 16 & 40 & A11 \\
      224 & 74 & 62 & 360 & 12 & 40 & A10 \\
      275 & 68 & 76 & 360 & 12 & 47 & A7 \\
      291 & 74 & 81 & 360 & 12 & 40 & A9 \\
    \end{tabular}
  \caption{%
    Parameters of different YBCO cJJs listed in the order of the increasing fitting parameter $I_\mathrm{c,0}$.
    Here, $I_\mathrm{c,0}$ denotes the depairing current at $T = 0\units{K}$ and $T_\mathrm{c}$ the critical temperature -- both obtained from the best fit of the data using equation~\eqref{eq:fiteq}.
    The current density $j_\mathrm{c,0}$ is calculated with $I_\mathrm{c,0}$ and the cross-section $A$ of the cJJ.
    $D_\mathrm{a}$ and $w$ are the fabrication parameters.
  }
  \label{tab:ict}
\end{table}

\Sec{Critical current \vs temperature}
\label{app:ict}

We have measured the $I_c(T)$ dependences for a set of cJJs embedded in the feed point of various microwave antennas, see Fig.~\ref{fig:antenna} as an example.
An overview of the fabrication parameters such as areal dose $D_\mathrm{a}$ and constriction width $w$, is presented in Tab.~\ref{tab:ict}.
The $I_\mathrm{c}(T)$ data, extracted from single IVCs at different temperatures with a voltage criterion of 50\units{\mu V} are presented in Fig.~\ref{fig:ict}.
Note that values of $I_\mathrm{c} < 0.3\units{\mu A}$ are not reliable due to the thermal noise at $T=4.2\units{K}$.
Therefore we adopted the cut-off criterion $I_\mathrm{cutoff}(T)= 0.3\units{\mu A}\cdot T\,/\, 4.2\units{K}$ and disregarded all points having $I_c(T)<I_\mathrm{cutoff}(T)$, see Fig~\ref{fig:ict}(a).
After such preprocessing, we tried to fit the data with the Bardeen equation \cite{bardeen1962}.
This approach, which describes the temperature dependence of the depairing current of a superconductor in the dirty limit was already used in several reports on cJJs \cite{uhl2024, desimoni2020}.
In short, the critical depairing current
\begin{equation}
  I_\mathrm{c}(T) = \frac{B_\mathrm{c}(T)}{\mu_0 \, \lambda_\mathrm{L}(T)} \approx I_\mathrm{c,0} \sbraces{1-\rbfrac{T}{T_c}^2}^\frac{3}{2},
  \label{eq:fiteq}
\end{equation}
where $I_\mathrm{c,0}$ is the depairing current at $T=0\units{K}$, $T_\mathrm{c}$ is the critical temperature.
The expression~\eqref{eq:fiteq} uses well known empirical dependencies for the bulk critical field $B_\mathrm{c}(T) = B_0 \sbraces{1 - (T/T_\mathrm{c})^2}$ and the London penetration depth $\lambda_\mathrm{L} = \lambda_0 / \sqrt{1 - (T/T_\mathrm{c})^4}$.
Finer details like d-wave order parameter symmetry in YBCO are ignored for now.

We used Eq.~\eqref{eq:fiteq} to fit our data with $I_\mathrm{c,0}$ and $T_\mathrm{c}$ as fitting parameters.
The fit boundaries for $T_\mathrm{c}$ were set as $T_c = 30$--$87\units{K}$.
A representive subset of devices compared with the fitting model is shown in Fig.~\ref{fig:ict}(b).
For lower critical currents, the noise contribution impacts huge parts of the acquired data.
Truncating the data by applying the cut-off criteria for the critical current, the fitting procedure is limited to a narrow range of critical currents, limiting its validity.
For higher critical currents the cut-off criteria barely shrinks the data.
Nevertheless, the Bardeen model accurately captures the temperature dependence of the critical current over a broad temperature range.

We point out, that the nominal width of the patterned cJJs may not necessarily represent the effective width of the device.
Due to possible fluctuations of the beam intensity and quality in addition to inhomogeinity in the film quality, we get $I_\mathrm{c}$ values which do not scale with the change in width and dose, cf Tab.~\ref{tab:ict}.

The depairing current density $j_\mathrm{c,0}$ can exceed \cite{arpaia2013, lang2005} $100\units{MA/cm^2}$.
As shown in Tab.~\ref{tab:ict}, the $j_\mathrm{c,0}$ values exhibit a considerable spread with the highest one being $81\units{MA/cm^2}$.
The much lower values of $j_\mathrm{c,0}$ may be due to an effective junction cross-section, which is lower than the nominal one.
Alternatively, the smaller bridges may show some degradation which may lead to a suppression of the order parameter, and thus an increased value of $\lambda_\mathrm{L}$ and a reduced value of $B_\mathrm{c}$.

In conclusion, the $I_c(T)$ behaviour is well described by the Bardeen expression~\eqref{eq:fiteq} for the depairing current vs $T$.
For higher $I_\mathrm{c,0}$ values, the devices can be used in applications at $50$--$70\units{K}$.

\bibliography{this}

\begin{thebibliography}{62}%
\makeatletter
\providecommand \@ifxundefined [1]{%
 \@ifx{#1\undefined}
}%
\providecommand \@ifnum [1]{%
 \ifnum #1\expandafter \@firstoftwo
 \else \expandafter \@secondoftwo
 \fi
}%
\providecommand \@ifx [1]{%
 \ifx #1\expandafter \@firstoftwo
 \else \expandafter \@secondoftwo
 \fi
}%
\providecommand \natexlab [1]{#1}%
\providecommand \enquote  [1]{``#1''}%
\providecommand \bibnamefont  [1]{#1}%
\providecommand \bibfnamefont [1]{#1}%
\providecommand \citenamefont [1]{#1}%
\providecommand \href@noop [0]{\@secondoftwo}%
\providecommand \href [0]{\begingroup \@sanitize@url \@href}%
\providecommand \@href[1]{\@@startlink{#1}\@@href}%
\providecommand \@@href[1]{\endgroup#1\@@endlink}%
\providecommand \@sanitize@url [0]{\catcode `\\12\catcode `\$12\catcode
  `\&12\catcode `\#12\catcode `\^12\catcode `\_12\catcode `\%12\relax}%
\providecommand \@@startlink[1]{}%
\providecommand \@@endlink[0]{}%
\providecommand \url  [0]{\begingroup\@sanitize@url \@url }%
\providecommand \@url [1]{\endgroup\@href {#1}{\urlprefix }}%
\providecommand \urlprefix  [0]{URL }%
\providecommand \Eprint [0]{\href }%
\providecommand \doibase [0]{http://dx.doi.org/}%
\providecommand \selectlanguage [0]{\@gobble}%
\providecommand \bibinfo  [0]{\@secondoftwo}%
\providecommand \bibfield  [0]{\@secondoftwo}%
\providecommand \translation [1]{[#1]}%
\providecommand \BibitemOpen [0]{}%
\providecommand \bibitemStop [0]{}%
\providecommand \bibitemNoStop [0]{.\EOS\space}%
\providecommand \EOS [0]{\spacefactor3000\relax}%
\providecommand \BibitemShut  [1]{\csname bibitem#1\endcsname}%
\let\auto@bib@innerbib\@empty
\bibitem [{\citenamefont {Vijay}\ \emph {et~al.}(2010)\citenamefont {Vijay},
  \citenamefont {{Levenson-Falk}}, \citenamefont {Slichter},\ and\
  \citenamefont {Siddiqi}}]{vijay2010}%
  \BibitemOpen
  \bibfield  {author} {\bibinfo {author} {\bibfnamefont {R.}~\bibnamefont
  {Vijay}}, \bibinfo {author} {\bibfnamefont {E.~M.}\ \bibnamefont
  {{Levenson-Falk}}}, \bibinfo {author} {\bibfnamefont {D.~H.}\ \bibnamefont
  {Slichter}}, \ and\ \bibinfo {author} {\bibfnamefont {I.}~\bibnamefont
  {Siddiqi}},\ }\bibfield  {title} {\enquote {\bibinfo {title} {Approaching
  ideal weak link behavior with three dimensional aluminum nanobridges},}\
  }\href {\doibase 10.1063/1.3443716} {\bibfield  {journal} {\bibinfo
  {journal} {Appl. Phys. Lett.}\ }\textbf {\bibinfo {volume} {96}},\ \bibinfo
  {pages} {223112} (\bibinfo {year} {2010})}\BibitemShut {NoStop}%
\bibitem [{\citenamefont {Kennedy}\ \emph {et~al.}(2019)\citenamefont
  {Kennedy}, \citenamefont {Burnett}, \citenamefont {Fenton}, \citenamefont
  {Constantino}, \citenamefont {Warburton}, \citenamefont {Morton},\ and\
  \citenamefont {{Dupont-Ferrier}}}]{kennedy2019}%
  \BibitemOpen
  \bibfield  {author} {\bibinfo {author} {\bibfnamefont {O.W.}\ \bibnamefont
  {Kennedy}}, \bibinfo {author} {\bibfnamefont {J.}~\bibnamefont {Burnett}},
  \bibinfo {author} {\bibfnamefont {J.C.}\ \bibnamefont {Fenton}}, \bibinfo
  {author} {\bibfnamefont {N.G.N.}\ \bibnamefont {Constantino}}, \bibinfo
  {author} {\bibfnamefont {P.A.}\ \bibnamefont {Warburton}}, \bibinfo {author}
  {\bibfnamefont {J.J.L.}\ \bibnamefont {Morton}}, \ and\ \bibinfo {author}
  {\bibfnamefont {E.}~\bibnamefont {{Dupont-Ferrier}}},\ }\bibfield  {title}
  {\enquote {\bibinfo {title} {Tunable {Nb} superconducting resonator based on
  a constriction nano-{SQUID} fabricated with a {Ne} focused ion beam},}\
  }\href {\doibase 10.1103/PhysRevApplied.11.014006} {\bibfield  {journal}
  {\bibinfo  {journal} {Phys. Rev. Appl.}\ }\textbf {\bibinfo {volume} {11}},\
  \bibinfo {pages} {014006} (\bibinfo {year} {2019})}\BibitemShut {NoStop}%
\bibitem [{\citenamefont {Uhl}\ \emph {et~al.}(2024)\citenamefont {Uhl},
  \citenamefont {Hackenbeck}, \citenamefont {Peter}, \citenamefont {Kleiner},
  \citenamefont {Koelle},\ and\ \citenamefont {Bothner}}]{uhl2024}%
  \BibitemOpen
  \bibfield  {author} {\bibinfo {author} {\bibfnamefont {Kevin}\ \bibnamefont
  {Uhl}}, \bibinfo {author} {\bibfnamefont {Daniel}\ \bibnamefont
  {Hackenbeck}}, \bibinfo {author} {\bibfnamefont {Janis}\ \bibnamefont
  {Peter}}, \bibinfo {author} {\bibfnamefont {Reinhold}\ \bibnamefont
  {Kleiner}}, \bibinfo {author} {\bibfnamefont {Dieter}\ \bibnamefont
  {Koelle}}, \ and\ \bibinfo {author} {\bibfnamefont {Daniel}\ \bibnamefont
  {Bothner}},\ }\bibfield  {title} {\enquote {\bibinfo {title} {Niobium quantum
  interference microwave circuits with monolithic three-dimensional nanobridge
  junctions},}\ }\href {\doibase 10.1103/PhysRevApplied.21.024051} {\bibfield
  {journal} {\bibinfo  {journal} {Phys. Rev. Appl.}\ }\textbf {\bibinfo
  {volume} {21}},\ \bibinfo {pages} {024051} (\bibinfo {year}
  {2024})}\BibitemShut {NoStop}%
\bibitem [{\citenamefont {Weber}\ \emph {et~al.}(2025)\citenamefont {Weber},
  \citenamefont {Jetter}, \citenamefont {Ullmann}, \citenamefont {Koch},
  \citenamefont {Pfander}, \citenamefont {Kress}, \citenamefont {Vervelaki},
  \citenamefont {Gross}, \citenamefont {Kieler}, \citenamefont {Drechsler},
  \citenamefont {Baral}, \citenamefont {Magrez}, \citenamefont {Kleiner},
  \citenamefont {Knoll}, \citenamefont {Poggio},\ and\ \citenamefont
  {Koelle}}]{Weber25}%
  \BibitemOpen
  \bibfield  {author} {\bibinfo {author} {\bibfnamefont {Timur}\ \bibnamefont
  {Weber}}, \bibinfo {author} {\bibfnamefont {Daniel}\ \bibnamefont {Jetter}},
  \bibinfo {author} {\bibfnamefont {Jan}\ \bibnamefont {Ullmann}}, \bibinfo
  {author} {\bibfnamefont {Simon~A.}\ \bibnamefont {Koch}}, \bibinfo {author}
  {\bibfnamefont {Simon~F.}\ \bibnamefont {Pfander}}, \bibinfo {author}
  {\bibfnamefont {Katharina}\ \bibnamefont {Kress}}, \bibinfo {author}
  {\bibfnamefont {Andriani}\ \bibnamefont {Vervelaki}}, \bibinfo {author}
  {\bibfnamefont {Boris}\ \bibnamefont {Gross}}, \bibinfo {author}
  {\bibfnamefont {Oliver}\ \bibnamefont {Kieler}}, \bibinfo {author}
  {\bibfnamefont {Ute}\ \bibnamefont {Drechsler}}, \bibinfo {author}
  {\bibfnamefont {Priya~R.}\ \bibnamefont {Baral}}, \bibinfo {author}
  {\bibfnamefont {Arnaud}\ \bibnamefont {Magrez}}, \bibinfo {author}
  {\bibfnamefont {Reinhold}\ \bibnamefont {Kleiner}}, \bibinfo {author}
  {\bibfnamefont {Armin~W.}\ \bibnamefont {Knoll}}, \bibinfo {author}
  {\bibfnamefont {Martino}\ \bibnamefont {Poggio}}, \ and\ \bibinfo {author}
  {\bibfnamefont {Dieter}\ \bibnamefont {Koelle}},\ }\bibfield  {title}
  {\enquote {\bibinfo {title} {Advanced {SQUID}-on-lever scanning probe for
  high-sensitivity magnetic microscopy with sub-100-nm spatial resolution},}\
  }\href {\doibase 10.1103/6s24-vz3k} {\bibfield  {journal} {\bibinfo
  {journal} {Phys. Rev. Appl}\ }\textbf {\bibinfo {volume} {24}},\ \bibinfo
  {pages} {054041} (\bibinfo {year} {2025})}\BibitemShut {NoStop}%
\bibitem [{\citenamefont {Granata}\ and\ \citenamefont
  {Vettoliere}(2016)}]{granata2016}%
  \BibitemOpen
  \bibfield  {author} {\bibinfo {author} {\bibfnamefont {Carmine}\ \bibnamefont
  {Granata}}\ and\ \bibinfo {author} {\bibfnamefont {Antonio}\ \bibnamefont
  {Vettoliere}},\ }\bibfield  {title} {\enquote {\bibinfo {title} {Nano
  {{Superconducting Quantum Interference}} device: {{A}} powerful tool for
  nanoscale investigations},}\ }\href {\doibase 10.1016/j.physrep.2015.12.001}
  {\bibfield  {journal} {\bibinfo  {journal} {Phys. Rep.}\ }\textbf {\bibinfo
  {volume} {614}},\ \bibinfo {pages} {1--69} (\bibinfo {year}
  {2016})}\BibitemShut {NoStop}%
\bibitem [{\citenamefont {Mart\'{i}nez-P\'{e}rez}\ and\ \citenamefont
  {Koelle}(2017)}]{Martinez-Perez17a}%
  \BibitemOpen
  \bibfield  {author} {\bibinfo {author} {\bibfnamefont {M.~J.}\ \bibnamefont
  {Mart\'{i}nez-P\'{e}rez}}\ and\ \bibinfo {author} {\bibfnamefont
  {D.}~\bibnamefont {Koelle}},\ }\bibfield  {title} {\enquote {\bibinfo {title}
  {{NanoSQUDs}: {B}asics \& recent advances},}\ }\href {\doibase
  10.1515/psr-2017-5001} {\bibfield  {journal} {\bibinfo  {journal} {Phys. Sci.
  Rev.}\ }\textbf {\bibinfo {volume} {2}},\ \bibinfo {pages} {20175001}
  (\bibinfo {year} {2017})}\BibitemShut {NoStop}%
\bibitem [{\citenamefont {Lam}\ and\ \citenamefont {Tilbrook}(2003)}]{lam2003}%
  \BibitemOpen
  \bibfield  {author} {\bibinfo {author} {\bibfnamefont {S.~K.~H.}\
  \bibnamefont {Lam}}\ and\ \bibinfo {author} {\bibfnamefont {D.~L.}\
  \bibnamefont {Tilbrook}},\ }\bibfield  {title} {\enquote {\bibinfo {title}
  {Development of a niobium nanosuperconducting quantum interference device for
  the detection of small spin populations},}\ }\href {\doibase
  10.1063/1.1554770} {\bibfield  {journal} {\bibinfo  {journal} {Appl. Phys.
  Lett.}\ }\textbf {\bibinfo {volume} {82}},\ \bibinfo {pages} {1078--1080}
  (\bibinfo {year} {2003})}\BibitemShut {NoStop}%
\bibitem [{\citenamefont {Wyss}\ \emph {et~al.}(2022)\citenamefont {Wyss},
  \citenamefont {Bagani}, \citenamefont {Jetter}, \citenamefont {Marchiori},
  \citenamefont {Vervelaki}, \citenamefont {Gross}, \citenamefont {Ridderbos},
  \citenamefont {Gliga}, \citenamefont {Sch{\"o}nenberger},\ and\ \citenamefont
  {Poggio}}]{wyss2022}%
  \BibitemOpen
  \bibfield  {author} {\bibinfo {author} {\bibfnamefont {M.}~\bibnamefont
  {Wyss}}, \bibinfo {author} {\bibfnamefont {K.}~\bibnamefont {Bagani}},
  \bibinfo {author} {\bibfnamefont {D.}~\bibnamefont {Jetter}}, \bibinfo
  {author} {\bibfnamefont {E.}~\bibnamefont {Marchiori}}, \bibinfo {author}
  {\bibfnamefont {A.}~\bibnamefont {Vervelaki}}, \bibinfo {author}
  {\bibfnamefont {B.}~\bibnamefont {Gross}}, \bibinfo {author} {\bibfnamefont
  {J.}~\bibnamefont {Ridderbos}}, \bibinfo {author} {\bibfnamefont
  {S.}~\bibnamefont {Gliga}}, \bibinfo {author} {\bibfnamefont
  {C.}~\bibnamefont {Sch{\"o}nenberger}}, \ and\ \bibinfo {author}
  {\bibfnamefont {M.}~\bibnamefont {Poggio}},\ }\bibfield  {title} {\enquote
  {\bibinfo {title} {Magnetic, {{Thermal}}, and {{Topographic Imaging}} with a
  {{Nanometer-Scale SQUID-On-Lever Scanning Probe}}},}\ }\href {\doibase
  10.1103/PhysRevApplied.17.034002} {\bibfield  {journal} {\bibinfo  {journal}
  {Phys. Rev. Appl.}\ }\textbf {\bibinfo {volume} {17}},\ \bibinfo {pages}
  {034002} (\bibinfo {year} {2022})}\BibitemShut {NoStop}%
\bibitem [{\citenamefont {Vasyukov}\ \emph {et~al.}(2013)\citenamefont
  {Vasyukov}, \citenamefont {Anahory}, \citenamefont {Embon}, \citenamefont
  {Halbertal}, \citenamefont {Cuppens}, \citenamefont {Neeman}, \citenamefont
  {Finkler}, \citenamefont {Segev}, \citenamefont {Myasoedov}, \citenamefont
  {Rappaport}, \citenamefont {Huber},\ and\ \citenamefont
  {Zeldov}}]{vasyukov2013}%
  \BibitemOpen
  \bibfield  {author} {\bibinfo {author} {\bibfnamefont {Denis}\ \bibnamefont
  {Vasyukov}}, \bibinfo {author} {\bibfnamefont {Yonathan}\ \bibnamefont
  {Anahory}}, \bibinfo {author} {\bibfnamefont {Lior}\ \bibnamefont {Embon}},
  \bibinfo {author} {\bibfnamefont {Dorri}\ \bibnamefont {Halbertal}}, \bibinfo
  {author} {\bibfnamefont {Jo}~\bibnamefont {Cuppens}}, \bibinfo {author}
  {\bibfnamefont {Lior}\ \bibnamefont {Neeman}}, \bibinfo {author}
  {\bibfnamefont {Amit}\ \bibnamefont {Finkler}}, \bibinfo {author}
  {\bibfnamefont {Yehonathan}\ \bibnamefont {Segev}}, \bibinfo {author}
  {\bibfnamefont {Yuri}\ \bibnamefont {Myasoedov}}, \bibinfo {author}
  {\bibfnamefont {Michael~L.}\ \bibnamefont {Rappaport}}, \bibinfo {author}
  {\bibfnamefont {Martin~E.}\ \bibnamefont {Huber}}, \ and\ \bibinfo {author}
  {\bibfnamefont {Eli}\ \bibnamefont {Zeldov}},\ }\bibfield  {title} {\enquote
  {\bibinfo {title} {A scanning superconducting quantum interference device
  with single electron spin sensitivity},}\ }\href {\doibase
  10.1038/nnano.2013.169} {\bibfield  {journal} {\bibinfo  {journal} {Nat.
  Nanotechnol.}\ }\textbf {\bibinfo {volume} {8}},\ \bibinfo {pages} {639--644}
  (\bibinfo {year} {2013})}\BibitemShut {NoStop}%
\bibitem [{\citenamefont {Anahory}\ \emph {et~al.}(2020)\citenamefont
  {Anahory}, \citenamefont {Naren}, \citenamefont {Lachman}, \citenamefont
  {Sinai}, \citenamefont {Uri}, \citenamefont {Embon}, \citenamefont {Yaakobi},
  \citenamefont {Myasoedov}, \citenamefont {Huber}, \citenamefont {Klajn},\
  and\ \citenamefont {Zeldov}}]{anahory2020}%
  \BibitemOpen
  \bibfield  {author} {\bibinfo {author} {\bibfnamefont {Y.}~\bibnamefont
  {Anahory}}, \bibinfo {author} {\bibfnamefont {H.~R.}\ \bibnamefont {Naren}},
  \bibinfo {author} {\bibfnamefont {E.~O.}\ \bibnamefont {Lachman}}, \bibinfo
  {author} {\bibfnamefont {S.~Buhbut}\ \bibnamefont {Sinai}}, \bibinfo {author}
  {\bibfnamefont {A.}~\bibnamefont {Uri}}, \bibinfo {author} {\bibfnamefont
  {L.}~\bibnamefont {Embon}}, \bibinfo {author} {\bibfnamefont
  {E.}~\bibnamefont {Yaakobi}}, \bibinfo {author} {\bibfnamefont
  {Y.}~\bibnamefont {Myasoedov}}, \bibinfo {author} {\bibfnamefont {M.~E.}\
  \bibnamefont {Huber}}, \bibinfo {author} {\bibfnamefont {R.}~\bibnamefont
  {Klajn}}, \ and\ \bibinfo {author} {\bibfnamefont {E.}~\bibnamefont
  {Zeldov}},\ }\bibfield  {title} {\enquote {\bibinfo {title} {{SQUID-on-tip}
  with single-electron spin sensitivity for high-field and ultra-low
  temperature nanomagnetic imaging},}\ }\href {\doibase 10.1039/C9NR08578E}
  {\bibfield  {journal} {\bibinfo  {journal} {Nanoscale}\ }\textbf {\bibinfo
  {volume} {12}},\ \bibinfo {pages} {3174--3182} (\bibinfo {year}
  {2020})}\BibitemShut {NoStop}%
\bibitem [{\citenamefont {Bothner}\ \emph {et~al.}(2021)\citenamefont
  {Bothner}, \citenamefont {Rodrigues},\ and\ \citenamefont
  {Steele}}]{bothner2021}%
  \BibitemOpen
  \bibfield  {author} {\bibinfo {author} {\bibfnamefont {D.}~\bibnamefont
  {Bothner}}, \bibinfo {author} {\bibfnamefont {I.~C.}\ \bibnamefont
  {Rodrigues}}, \ and\ \bibinfo {author} {\bibfnamefont {G.~A.}\ \bibnamefont
  {Steele}},\ }\bibfield  {title} {\enquote {\bibinfo {title} {Photon-pressure
  strong coupling between two superconducting circuits},}\ }\href {\doibase
  10.1038/s41567-020-0987-5} {\bibfield  {journal} {\bibinfo  {journal} {Nat.
  Phys.}\ }\textbf {\bibinfo {volume} {17}},\ \bibinfo {pages} {85--91}
  (\bibinfo {year} {2021})}\BibitemShut {NoStop}%
\bibitem [{\citenamefont {Rodrigues}\ \emph {et~al.}(2021)\citenamefont
  {Rodrigues}, \citenamefont {Bothner},\ and\ \citenamefont
  {Steele}}]{rodrigues2021}%
  \BibitemOpen
  \bibfield  {author} {\bibinfo {author} {\bibfnamefont {Ines~Corveira}\
  \bibnamefont {Rodrigues}}, \bibinfo {author} {\bibfnamefont {Daniel}\
  \bibnamefont {Bothner}}, \ and\ \bibinfo {author} {\bibfnamefont
  {Gary~Alexander}\ \bibnamefont {Steele}},\ }\bibfield  {title} {\enquote
  {\bibinfo {title} {Cooling photon-pressure circuits into the quantum
  regime},}\ }\href {\doibase 10.1126/sciadv.abg6653} {\bibfield  {journal}
  {\bibinfo  {journal} {Sci. Adv.}\ }\textbf {\bibinfo {volume} {7}},\ \bibinfo
  {pages} {eabg6653} (\bibinfo {year} {2021})}\BibitemShut {NoStop}%
\bibitem [{\citenamefont {Rieger}\ \emph {et~al.}(2023)\citenamefont {Rieger},
  \citenamefont {G{\"u}nzler}, \citenamefont {Spiecker}, \citenamefont
  {Paluch}, \citenamefont {Winkel}, \citenamefont {Hahn}, \citenamefont
  {Hohmann}, \citenamefont {Bacher}, \citenamefont {Wernsdorfer},\ and\
  \citenamefont {Pop}}]{rieger2023}%
  \BibitemOpen
  \bibfield  {author} {\bibinfo {author} {\bibfnamefont {D.}~\bibnamefont
  {Rieger}}, \bibinfo {author} {\bibfnamefont {S.}~\bibnamefont {G{\"u}nzler}},
  \bibinfo {author} {\bibfnamefont {M.}~\bibnamefont {Spiecker}}, \bibinfo
  {author} {\bibfnamefont {P.}~\bibnamefont {Paluch}}, \bibinfo {author}
  {\bibfnamefont {P.}~\bibnamefont {Winkel}}, \bibinfo {author} {\bibfnamefont
  {L.}~\bibnamefont {Hahn}}, \bibinfo {author} {\bibfnamefont {J.~K.}\
  \bibnamefont {Hohmann}}, \bibinfo {author} {\bibfnamefont {A.}~\bibnamefont
  {Bacher}}, \bibinfo {author} {\bibfnamefont {W.}~\bibnamefont {Wernsdorfer}},
  \ and\ \bibinfo {author} {\bibfnamefont {I.~M.}\ \bibnamefont {Pop}},\
  }\bibfield  {title} {\enquote {\bibinfo {title} {Granular aluminium
  nanojunction fluxonium qubit},}\ }\href {\doibase 10.1038/s41563-022-01417-9}
  {\bibfield  {journal} {\bibinfo  {journal} {Nat. Mater.}\ }\textbf {\bibinfo
  {volume} {22}},\ \bibinfo {pages} {194--199} (\bibinfo {year}
  {2023})}\BibitemShut {NoStop}%
\bibitem [{\citenamefont {Likharev}(1979)}]{likharev1979}%
  \BibitemOpen
  \bibfield  {author} {\bibinfo {author} {\bibfnamefont {K.~K.}\ \bibnamefont
  {Likharev}},\ }\bibfield  {title} {\enquote {\bibinfo {title}
  {Superconducting weak links},}\ }\href {\doibase 10.1103/RevModPhys.51.101}
  {\bibfield  {journal} {\bibinfo  {journal} {Rev. Mod. Phys.}\ }\textbf
  {\bibinfo {volume} {51}},\ \bibinfo {pages} {101--159} (\bibinfo {year}
  {1979})}\BibitemShut {NoStop}%
\bibitem [{\citenamefont {Golubov}\ \emph {et~al.}(2004)\citenamefont
  {Golubov}, \citenamefont {Kupriyanov},\ and\ \citenamefont
  {Il'ichev}}]{golubov2004a}%
  \BibitemOpen
  \bibfield  {author} {\bibinfo {author} {\bibfnamefont {A.~A.}\ \bibnamefont
  {Golubov}}, \bibinfo {author} {\bibfnamefont {M.~{\relax Yu}.}\ \bibnamefont
  {Kupriyanov}}, \ and\ \bibinfo {author} {\bibfnamefont {E.}~\bibnamefont
  {Il'ichev}},\ }\bibfield  {title} {\enquote {\bibinfo {title} {The
  current-phase relation in {{Josephson}} junctions},}\ }\href {\doibase
  10.1103/RevModPhys.76.411} {\bibfield  {journal} {\bibinfo  {journal} {Rev.
  Mod. Phys.}\ }\textbf {\bibinfo {volume} {76}},\ \bibinfo {pages} {411--469}
  (\bibinfo {year} {2004})}\BibitemShut {NoStop}%
\bibitem [{\citenamefont {Van~Weerdenburg}\ \emph {et~al.}(2023)\citenamefont
  {Van~Weerdenburg}, \citenamefont {Kamlapure}, \citenamefont {Fyhn},
  \citenamefont {Huang}, \citenamefont {Van~Mullekom}, \citenamefont
  {Steinbrecher}, \citenamefont {Krogstrup}, \citenamefont {Linder},\ and\
  \citenamefont {Khajetoorians}}]{vanweerdenburg2023}%
  \BibitemOpen
  \bibfield  {author} {\bibinfo {author} {\bibfnamefont {Werner M.~J.}\
  \bibnamefont {Van~Weerdenburg}}, \bibinfo {author} {\bibfnamefont {Anand}\
  \bibnamefont {Kamlapure}}, \bibinfo {author} {\bibfnamefont {Eirik~Holm}\
  \bibnamefont {Fyhn}}, \bibinfo {author} {\bibfnamefont {Xiaochun}\
  \bibnamefont {Huang}}, \bibinfo {author} {\bibfnamefont {Niels P.~E.}\
  \bibnamefont {Van~Mullekom}}, \bibinfo {author} {\bibfnamefont {Manuel}\
  \bibnamefont {Steinbrecher}}, \bibinfo {author} {\bibfnamefont {Peter}\
  \bibnamefont {Krogstrup}}, \bibinfo {author} {\bibfnamefont {Jacob}\
  \bibnamefont {Linder}}, \ and\ \bibinfo {author} {\bibfnamefont
  {Alexander~Ako}\ \bibnamefont {Khajetoorians}},\ }\bibfield  {title}
  {\enquote {\bibinfo {title} {Extreme enhancement of superconductivity in
  epitaxial aluminum near the monolayer limit},}\ }\href {\doibase
  10.1126/sciadv.adf5500} {\bibfield  {journal} {\bibinfo  {journal} {Sci.
  Adv.}\ }\textbf {\bibinfo {volume} {9}},\ \bibinfo {pages} {eadf5500}
  (\bibinfo {year} {2023})}\BibitemShut {NoStop}%
\bibitem [{\citenamefont {{L{\'o}pez-N{\'u}{\~n}ez}}\ \emph
  {et~al.}(2025)\citenamefont {{L{\'o}pez-N{\'u}{\~n}ez}}, \citenamefont
  {{Torras-Coloma}}, \citenamefont {{Portell-Montserrat}}, \citenamefont
  {Bertoldo}, \citenamefont {Cozzolino}, \citenamefont {Ummarino},
  \citenamefont {Zaccone}, \citenamefont {Rius}, \citenamefont {Martinez},\
  and\ \citenamefont {{Forn-D{\'i}az}}}]{lopez-nunez2025}%
  \BibitemOpen
  \bibfield  {author} {\bibinfo {author} {\bibfnamefont {David}\ \bibnamefont
  {{L{\'o}pez-N{\'u}{\~n}ez}}}, \bibinfo {author} {\bibfnamefont {Alba}\
  \bibnamefont {{Torras-Coloma}}}, \bibinfo {author} {\bibfnamefont {Queralt}\
  \bibnamefont {{Portell-Montserrat}}}, \bibinfo {author} {\bibfnamefont
  {Elia}\ \bibnamefont {Bertoldo}}, \bibinfo {author} {\bibfnamefont {Luca}\
  \bibnamefont {Cozzolino}}, \bibinfo {author} {\bibfnamefont
  {Giovanni~Alberto}\ \bibnamefont {Ummarino}}, \bibinfo {author}
  {\bibfnamefont {Alessio}\ \bibnamefont {Zaccone}}, \bibinfo {author}
  {\bibfnamefont {Gemma}\ \bibnamefont {Rius}}, \bibinfo {author}
  {\bibfnamefont {Manel}\ \bibnamefont {Martinez}}, \ and\ \bibinfo {author}
  {\bibfnamefont {Pol}\ \bibnamefont {{Forn-D{\'i}az}}},\ }\bibfield  {title}
  {\enquote {\bibinfo {title} {Superconducting penetration depth of
  {{Aluminum}} thin films},}\ }\href {\doibase 10.1088/1361-6668/adf360}
  {\bibfield  {journal} {\bibinfo  {journal} {Supercond. Sci. Technol.}\
  }\textbf {\bibinfo {volume} {38}},\ \bibinfo {pages} {095004} (\bibinfo
  {year} {2025})}\BibitemShut {NoStop}%
\bibitem [{\citenamefont {Pinto}\ \emph {et~al.}(2018)\citenamefont {Pinto},
  \citenamefont {Rezvani}, \citenamefont {Perali}, \citenamefont {Flammia},
  \citenamefont {Milo{\v s}evi{\'c}}, \citenamefont {Fretto}, \citenamefont
  {Cassiago},\ and\ \citenamefont {De~Leo}}]{pinto2018}%
  \BibitemOpen
  \bibfield  {author} {\bibinfo {author} {\bibfnamefont {Nicola}\ \bibnamefont
  {Pinto}}, \bibinfo {author} {\bibfnamefont {S.~Javad}\ \bibnamefont
  {Rezvani}}, \bibinfo {author} {\bibfnamefont {Andrea}\ \bibnamefont
  {Perali}}, \bibinfo {author} {\bibfnamefont {Luca}\ \bibnamefont {Flammia}},
  \bibinfo {author} {\bibfnamefont {Milorad~V.}\ \bibnamefont {Milo{\v
  s}evi{\'c}}}, \bibinfo {author} {\bibfnamefont {Matteo}\ \bibnamefont
  {Fretto}}, \bibinfo {author} {\bibfnamefont {Cristina}\ \bibnamefont
  {Cassiago}}, \ and\ \bibinfo {author} {\bibfnamefont {Natascia}\ \bibnamefont
  {De~Leo}},\ }\bibfield  {title} {\enquote {\bibinfo {title} {Dimensional
  crossover and incipient quantum size effects in superconducting niobium
  nanofilms},}\ }\href {\doibase 10.1038/s41598-018-22983-6} {\bibfield
  {journal} {\bibinfo  {journal} {Sci. Rep.}\ }\textbf {\bibinfo {volume}
  {8}},\ \bibinfo {pages} {4710} (\bibinfo {year} {2018})}\BibitemShut
  {NoStop}%
\bibitem [{\citenamefont {Schneider}\ \emph {et~al.}(1993)\citenamefont
  {Schneider}, \citenamefont {Kohlstedt},\ and\ \citenamefont
  {W{\"o}rdenweber}}]{schneider1993}%
  \BibitemOpen
  \bibfield  {author} {\bibinfo {author} {\bibfnamefont {J.}~\bibnamefont
  {Schneider}}, \bibinfo {author} {\bibfnamefont {H.}~\bibnamefont
  {Kohlstedt}}, \ and\ \bibinfo {author} {\bibfnamefont {R.}~\bibnamefont
  {W{\"o}rdenweber}},\ }\bibfield  {title} {\enquote {\bibinfo {title}
  {Nanobridges of optimized {YBa$_2$Cu$_3$O$_7$} thin films for superconducting
  flux-flow type devices},}\ }\href {\doibase 10.1063/1.110496} {\bibfield
  {journal} {\bibinfo  {journal} {Appl. Phys. Lett.}\ }\textbf {\bibinfo
  {volume} {63}},\ \bibinfo {pages} {2426--2428} (\bibinfo {year}
  {1993})}\BibitemShut {NoStop}%
\bibitem [{\citenamefont {Papari}\ \emph {et~al.}(2012)\citenamefont {Papari},
  \citenamefont {Carillo}, \citenamefont {Stornaiuolo}, \citenamefont
  {Longobardi}, \citenamefont {Beltram},\ and\ \citenamefont
  {Tafuri}}]{papari2012}%
  \BibitemOpen
  \bibfield  {author} {\bibinfo {author} {\bibfnamefont {G.}~\bibnamefont
  {Papari}}, \bibinfo {author} {\bibfnamefont {F.}~\bibnamefont {Carillo}},
  \bibinfo {author} {\bibfnamefont {D.}~\bibnamefont {Stornaiuolo}}, \bibinfo
  {author} {\bibfnamefont {L.}~\bibnamefont {Longobardi}}, \bibinfo {author}
  {\bibfnamefont {F.}~\bibnamefont {Beltram}}, \ and\ \bibinfo {author}
  {\bibfnamefont {F.}~\bibnamefont {Tafuri}},\ }\bibfield  {title} {\enquote
  {\bibinfo {title} {High critical current density and scaling of phase-slip
  processes in {{YBaCuO}} nanowires},}\ }\href {\doibase
  10.1088/0953-2048/25/3/035011} {\bibfield  {journal} {\bibinfo  {journal}
  {Supercond. Sci. Technol.}\ }\textbf {\bibinfo {volume} {25}},\ \bibinfo
  {pages} {035011} (\bibinfo {year} {2012})}\BibitemShut {NoStop}%
\bibitem [{\citenamefont {Larsson}\ \emph {et~al.}(2000)\citenamefont
  {Larsson}, \citenamefont {Nilsson},\ and\ \citenamefont
  {Ivanov}}]{Larsson2000}%
  \BibitemOpen
  \bibfield  {author} {\bibinfo {author} {\bibfnamefont {P.}~\bibnamefont
  {Larsson}}, \bibinfo {author} {\bibfnamefont {B.}~\bibnamefont {Nilsson}}, \
  and\ \bibinfo {author} {\bibfnamefont {Z.~G.}\ \bibnamefont {Ivanov}},\
  }\bibfield  {title} {\enquote {\bibinfo {title} {Fabrication and transport
  measurements of {YBa$_2$Cu$_3$O$_{7-x}$} nanostructures},}\ }\href {\doibase
  10.1116/1.591145} {\bibfield  {journal} {\bibinfo  {journal} {J. Vac. Sci.
  Technol. B}\ }\textbf {\bibinfo {volume} {18}},\ \bibinfo {pages} {25--31}
  (\bibinfo {year} {2000})}\BibitemShut {NoStop}%
\bibitem [{\citenamefont {Tafuri}\ \emph {et~al.}(2013)\citenamefont {Tafuri},
  \citenamefont {Massarotti}, \citenamefont {Galletti}, \citenamefont
  {Stornaiuolo}, \citenamefont {Montemurro}, \citenamefont {Longobardi},
  \citenamefont {Lucignano}, \citenamefont {Rotoli}, \citenamefont {Pepe},
  \citenamefont {Tagliacozzo},\ and\ \citenamefont {Lombardi}}]{tafuri2013}%
  \BibitemOpen
  \bibfield  {author} {\bibinfo {author} {\bibfnamefont {Francesco}\
  \bibnamefont {Tafuri}}, \bibinfo {author} {\bibfnamefont {Davide}\
  \bibnamefont {Massarotti}}, \bibinfo {author} {\bibfnamefont {Luca}\
  \bibnamefont {Galletti}}, \bibinfo {author} {\bibfnamefont {Daniela}\
  \bibnamefont {Stornaiuolo}}, \bibinfo {author} {\bibfnamefont {Domenico}\
  \bibnamefont {Montemurro}}, \bibinfo {author} {\bibfnamefont {Luigi}\
  \bibnamefont {Longobardi}}, \bibinfo {author} {\bibfnamefont {Procolo}\
  \bibnamefont {Lucignano}}, \bibinfo {author} {\bibfnamefont {Giacomo}\
  \bibnamefont {Rotoli}}, \bibinfo {author} {\bibfnamefont {Giovanni~Piero}\
  \bibnamefont {Pepe}}, \bibinfo {author} {\bibfnamefont {Arturo}\ \bibnamefont
  {Tagliacozzo}}, \ and\ \bibinfo {author} {\bibfnamefont {Floriana}\
  \bibnamefont {Lombardi}},\ }\bibfield  {title} {\enquote {\bibinfo {title}
  {Recent {Achievements} on the {Physics} of {High-$T_C$ Superconductor
  Josephson Junctions}: {Background}, {Perspectives} and {Inspiration}},}\
  }\href {\doibase 10.1007/s10948-012-1773-0} {\bibfield  {journal} {\bibinfo
  {journal} {J. Supercond. Nov. Magn.}\ }\textbf {\bibinfo {volume} {26}},\
  \bibinfo {pages} {21--41} (\bibinfo {year} {2013})}\BibitemShut {NoStop}%
\bibitem [{\citenamefont {Nawaz}\ \emph
  {et~al.}(2013{\natexlab{a}})\citenamefont {Nawaz}, \citenamefont {Arpaia},
  \citenamefont {Bauch},\ and\ \citenamefont {Lombardi}}]{nawaz2013}%
  \BibitemOpen
  \bibfield  {author} {\bibinfo {author} {\bibfnamefont {S.}~\bibnamefont
  {Nawaz}}, \bibinfo {author} {\bibfnamefont {R.}~\bibnamefont {Arpaia}},
  \bibinfo {author} {\bibfnamefont {T.}~\bibnamefont {Bauch}}, \ and\ \bibinfo
  {author} {\bibfnamefont {F.}~\bibnamefont {Lombardi}},\ }\bibfield  {title}
  {\enquote {\bibinfo {title} {Approaching the theoretical depairing current in
  {YBa$_2$Cu$_3$O$_{7-x}$} nanowires},}\ }\href {\doibase
  10.1016/j.physc.2013.07.011} {\bibfield  {journal} {\bibinfo  {journal}
  {Physica C}\ }\textbf {\bibinfo {volume} {495}},\ \bibinfo {pages} {33--38}
  (\bibinfo {year} {2013}{\natexlab{a}})}\BibitemShut {NoStop}%
\bibitem [{\citenamefont {Arpaia}\ \emph {et~al.}(2013)\citenamefont {Arpaia},
  \citenamefont {Nawaz}, \citenamefont {Lombardi},\ and\ \citenamefont
  {Bauch}}]{arpaia2013}%
  \BibitemOpen
  \bibfield  {author} {\bibinfo {author} {\bibfnamefont {R.}~\bibnamefont
  {Arpaia}}, \bibinfo {author} {\bibfnamefont {S.}~\bibnamefont {Nawaz}},
  \bibinfo {author} {\bibfnamefont {F.}~\bibnamefont {Lombardi}}, \ and\
  \bibinfo {author} {\bibfnamefont {T.}~\bibnamefont {Bauch}},\ }\bibfield
  {title} {\enquote {\bibinfo {title} {Improved nanopatterning for {YBCO}
  nanowires approaching the depairing current},}\ }\href {\doibase
  10.1109/TASC.2013.2247454} {\bibfield  {journal} {\bibinfo  {journal} {IEEE
  Trans. Appl. Supercond.}\ }\textbf {\bibinfo {volume} {23}},\ \bibinfo
  {pages} {1101505--1101505} (\bibinfo {year} {2013})}\BibitemShut {NoStop}%
\bibitem [{\citenamefont {Arpaia}\ \emph {et~al.}(2016)\citenamefont {Arpaia},
  \citenamefont {Arzeo}, \citenamefont {Baghdadi}, \citenamefont {Trabaldo},
  \citenamefont {Lombardi},\ and\ \citenamefont {Bauch}}]{arpaia2016}%
  \BibitemOpen
  \bibfield  {author} {\bibinfo {author} {\bibfnamefont {R.}~\bibnamefont
  {Arpaia}}, \bibinfo {author} {\bibfnamefont {M.}~\bibnamefont {Arzeo}},
  \bibinfo {author} {\bibfnamefont {R.}~\bibnamefont {Baghdadi}}, \bibinfo
  {author} {\bibfnamefont {E.}~\bibnamefont {Trabaldo}}, \bibinfo {author}
  {\bibfnamefont {F.}~\bibnamefont {Lombardi}}, \ and\ \bibinfo {author}
  {\bibfnamefont {T.}~\bibnamefont {Bauch}},\ }\bibfield  {title} {\enquote
  {\bibinfo {title} {Improved noise performance of ultrathin {{YBCO Dayem}}
  bridge {{nanoSQUIDs}}},}\ }\href {\doibase 10.1088/0953-2048/30/1/014008}
  {\bibfield  {journal} {\bibinfo  {journal} {Supercond. Sci. Technol.}\
  }\textbf {\bibinfo {volume} {30}},\ \bibinfo {pages} {014008} (\bibinfo
  {year} {2016})}\BibitemShut {NoStop}%
\bibitem [{\citenamefont {Arpaia}\ \emph {et~al.}(2017)\citenamefont {Arpaia},
  \citenamefont {Golubev}, \citenamefont {Baghdadi}, \citenamefont {Ciancio},
  \citenamefont {Dra{\v z}i{\'c}}, \citenamefont {Orgiani}, \citenamefont
  {Montemurro}, \citenamefont {Bauch},\ and\ \citenamefont
  {Lombardi}}]{arpaia2017}%
  \BibitemOpen
  \bibfield  {author} {\bibinfo {author} {\bibfnamefont {Riccardo}\
  \bibnamefont {Arpaia}}, \bibinfo {author} {\bibfnamefont {Dmitri}\
  \bibnamefont {Golubev}}, \bibinfo {author} {\bibfnamefont {Reza}\
  \bibnamefont {Baghdadi}}, \bibinfo {author} {\bibfnamefont {Regina}\
  \bibnamefont {Ciancio}}, \bibinfo {author} {\bibfnamefont {Goran}\
  \bibnamefont {Dra{\v z}i{\'c}}}, \bibinfo {author} {\bibfnamefont {Pasquale}\
  \bibnamefont {Orgiani}}, \bibinfo {author} {\bibfnamefont {Domenico}\
  \bibnamefont {Montemurro}}, \bibinfo {author} {\bibfnamefont {Thilo}\
  \bibnamefont {Bauch}}, \ and\ \bibinfo {author} {\bibfnamefont {Floriana}\
  \bibnamefont {Lombardi}},\ }\bibfield  {title} {\enquote {\bibinfo {title}
  {Transport properties of ultrathin {YBa$_2$Cu$_3$O$_{7-\delta}$} nanowires:
  {{A}} route to single-photon detection},}\ }\href {\doibase
  10.1103/PhysRevB.96.064525} {\bibfield  {journal} {\bibinfo  {journal} {Phys.
  Rev. B}\ }\textbf {\bibinfo {volume} {96}},\ \bibinfo {pages} {064525}
  (\bibinfo {year} {2017})}\BibitemShut {NoStop}%
\bibitem [{\citenamefont {Nawaz}\ \emph
  {et~al.}(2013{\natexlab{b}})\citenamefont {Nawaz}, \citenamefont {Arpaia},
  \citenamefont {Lombardi},\ and\ \citenamefont {Bauch}}]{nawaz2013a}%
  \BibitemOpen
  \bibfield  {author} {\bibinfo {author} {\bibfnamefont {S.}~\bibnamefont
  {Nawaz}}, \bibinfo {author} {\bibfnamefont {R.}~\bibnamefont {Arpaia}},
  \bibinfo {author} {\bibfnamefont {F.}~\bibnamefont {Lombardi}}, \ and\
  \bibinfo {author} {\bibfnamefont {T.}~\bibnamefont {Bauch}},\ }\bibfield
  {title} {\enquote {\bibinfo {title} {Microwave response of superconducting
  {YBa$_2$Cu$_3$O$_{7-\delta}$} nanowire bridges sustaining the critical
  depairing current: {Evidence} of {Josephson-like Behavior}},}\ }\href
  {\doibase 10.1103/PhysRevLett.110.167004} {\bibfield  {journal} {\bibinfo
  {journal} {Phys. Rev. Lett.}\ }\textbf {\bibinfo {volume} {110}},\ \bibinfo
  {pages} {167004} (\bibinfo {year} {2013}{\natexlab{b}})}\BibitemShut
  {NoStop}%
\bibitem [{\citenamefont {Trabaldo}\ \emph {et~al.}(2019)\citenamefont
  {Trabaldo}, \citenamefont {Arpaia}, \citenamefont {Arzeo}, \citenamefont
  {Andersson}, \citenamefont {Golubev}, \citenamefont {Lombardi},\ and\
  \citenamefont {Bauch}}]{trabaldo2019}%
  \BibitemOpen
  \bibfield  {author} {\bibinfo {author} {\bibfnamefont {E}~\bibnamefont
  {Trabaldo}}, \bibinfo {author} {\bibfnamefont {R}~\bibnamefont {Arpaia}},
  \bibinfo {author} {\bibfnamefont {M}~\bibnamefont {Arzeo}}, \bibinfo {author}
  {\bibfnamefont {E}~\bibnamefont {Andersson}}, \bibinfo {author}
  {\bibfnamefont {D}~\bibnamefont {Golubev}}, \bibinfo {author} {\bibfnamefont
  {F}~\bibnamefont {Lombardi}}, \ and\ \bibinfo {author} {\bibfnamefont
  {T}~\bibnamefont {Bauch}},\ }\bibfield  {title} {\enquote {\bibinfo {title}
  {Transport and noise properties of {{YBCO}} nanowire based {{nanoSQUIDs}}},}\
  }\href {\doibase 10.1088/1361-6668/ab1814} {\bibfield  {journal} {\bibinfo
  {journal} {Supercond. Sci. Technol.}\ }\textbf {\bibinfo {volume} {32}},\
  \bibinfo {pages} {073001} (\bibinfo {year} {2019})}\BibitemShut {NoStop}%
\bibitem [{\citenamefont {Trabaldo}\ \emph {et~al.}(2020)\citenamefont
  {Trabaldo}, \citenamefont {Ruffieux}, \citenamefont {Andersson},
  \citenamefont {Arpaia}, \citenamefont {Montemurro}, \citenamefont
  {Schneiderman}, \citenamefont {Kalaboukhov}, \citenamefont {Winkler},
  \citenamefont {Lombardi},\ and\ \citenamefont {Bauch}}]{trabaldo2020}%
  \BibitemOpen
  \bibfield  {author} {\bibinfo {author} {\bibfnamefont {E.}~\bibnamefont
  {Trabaldo}}, \bibinfo {author} {\bibfnamefont {S.}~\bibnamefont {Ruffieux}},
  \bibinfo {author} {\bibfnamefont {E.}~\bibnamefont {Andersson}}, \bibinfo
  {author} {\bibfnamefont {R.}~\bibnamefont {Arpaia}}, \bibinfo {author}
  {\bibfnamefont {D.}~\bibnamefont {Montemurro}}, \bibinfo {author}
  {\bibfnamefont {J.~F.}\ \bibnamefont {Schneiderman}}, \bibinfo {author}
  {\bibfnamefont {A.}~\bibnamefont {Kalaboukhov}}, \bibinfo {author}
  {\bibfnamefont {D.}~\bibnamefont {Winkler}}, \bibinfo {author} {\bibfnamefont
  {F.}~\bibnamefont {Lombardi}}, \ and\ \bibinfo {author} {\bibfnamefont
  {T.}~\bibnamefont {Bauch}},\ }\bibfield  {title} {\enquote {\bibinfo {title}
  {Properties of grooved {Dayem} bridge based {YBa$_2$Cu$_3$O$_{7-\delta}$}
  superconducting quantum interference devices and magnetometers},}\ }\href
  {\doibase 10.1063/5.0001805} {\bibfield  {journal} {\bibinfo  {journal}
  {Appl. Phys. Lett.}\ }\textbf {\bibinfo {volume} {116}},\ \bibinfo {pages}
  {132601} (\bibinfo {year} {2020})}\BibitemShut {NoStop}%
\bibitem [{\citenamefont {Ma}\ \emph {et~al.}(2025)\citenamefont {Ma},
  \citenamefont {Wang}, \citenamefont {Wang}, \citenamefont {Luo},
  \citenamefont {Li}, \citenamefont {Wang}, \citenamefont {Du}, \citenamefont
  {Yang}, \citenamefont {Huang}, \citenamefont {Wu}, \citenamefont {Li},
  \citenamefont {Wang}, \citenamefont {Liu},\ and\ \citenamefont
  {Li}}]{ma2025}%
  \BibitemOpen
  \bibfield  {author} {\bibinfo {author} {\bibfnamefont {Huiqin}\ \bibnamefont
  {Ma}}, \bibinfo {author} {\bibfnamefont {Hanbin}\ \bibnamefont {Wang}},
  \bibinfo {author} {\bibfnamefont {Yang}\ \bibnamefont {Wang}}, \bibinfo
  {author} {\bibfnamefont {Zhengyang}\ \bibnamefont {Luo}}, \bibinfo {author}
  {\bibfnamefont {Zongpei}\ \bibnamefont {Li}}, \bibinfo {author}
  {\bibfnamefont {Yong}\ \bibnamefont {Wang}}, \bibinfo {author} {\bibfnamefont
  {Xinchuan}\ \bibnamefont {Du}}, \bibinfo {author} {\bibfnamefont {Chao}\
  \bibnamefont {Yang}}, \bibinfo {author} {\bibfnamefont {Jianwen}\
  \bibnamefont {Huang}}, \bibinfo {author} {\bibfnamefont {Chunyang}\
  \bibnamefont {Wu}}, \bibinfo {author} {\bibfnamefont {Nannan}\ \bibnamefont
  {Li}}, \bibinfo {author} {\bibfnamefont {Xianfu}\ \bibnamefont {Wang}},
  \bibinfo {author} {\bibfnamefont {Yuqing}\ \bibnamefont {Liu}}, \ and\
  \bibinfo {author} {\bibfnamefont {Peng}\ \bibnamefont {Li}},\ }\bibfield
  {title} {\enquote {\bibinfo {title} {Minimal {Damage} and {Tunable
  Fabrication} of {Atomic}-{Scale Ultrathin YBa$_2$Cu$_3$O$_{7-\delta}$}
  {Nanowires} with {High Uniformity}},}\ }\href {\doibase
  10.1002/sstr.202400661} {\bibfield  {journal} {\bibinfo  {journal} {Small
  Struct.}\ }\textbf {\bibinfo {volume} {6}},\ \bibinfo {pages} {2400661}
  (\bibinfo {year} {2025})}\BibitemShut {NoStop}%
\bibitem [{\citenamefont {Xu}\ and\ \citenamefont {Heath}(2008)}]{xu2008}%
  \BibitemOpen
  \bibfield  {author} {\bibinfo {author} {\bibfnamefont {Ke}~\bibnamefont
  {Xu}}\ and\ \bibinfo {author} {\bibfnamefont {James~R.}\ \bibnamefont
  {Heath}},\ }\bibfield  {title} {\enquote {\bibinfo {title} {Long,
  {{Highly-Ordered High-Temperature Superconductor Nanowire Arrays}}},}\ }\href
  {\doibase 10.1021/nl802264x} {\bibfield  {journal} {\bibinfo  {journal} {Nano
  Lett.}\ }\textbf {\bibinfo {volume} {8}},\ \bibinfo {pages} {3845--3849}
  (\bibinfo {year} {2008})}\BibitemShut {NoStop}%
\bibitem [{\citenamefont {Lam}\ \emph {et~al.}(2019)\citenamefont {Lam},
  \citenamefont {Bendavid},\ and\ \citenamefont {Du}}]{lam2019}%
  \BibitemOpen
  \bibfield  {author} {\bibinfo {author} {\bibfnamefont {Simon K~H}\
  \bibnamefont {Lam}}, \bibinfo {author} {\bibfnamefont {Avi}\ \bibnamefont
  {Bendavid}}, \ and\ \bibinfo {author} {\bibfnamefont {Jia}\ \bibnamefont
  {Du}},\ }\bibfield  {title} {\enquote {\bibinfo {title} {Hot spot formation
  in focused-ion-beam-fabricated {YBa$_2$Cu$_3$O$_{7-x}$} nanobridges with high
  critical current densities},}\ }\href {\doibase 10.1088/1361-6528/ab1971}
  {\bibfield  {journal} {\bibinfo  {journal} {Nanotechnology}\ }\textbf
  {\bibinfo {volume} {30}},\ \bibinfo {pages} {325301} (\bibinfo {year}
  {2019})}\BibitemShut {NoStop}%
\bibitem [{\citenamefont {Lyatti}\ \emph {et~al.}(2018)\citenamefont {Lyatti},
  \citenamefont {Wolff}, \citenamefont {Savenko}, \citenamefont {Kruth},
  \citenamefont {Ferrari}, \citenamefont {Poppe}, \citenamefont {Pernice},
  \citenamefont {{Dunin-Borkowski}},\ and\ \citenamefont
  {Schuck}}]{lyatti2018}%
  \BibitemOpen
  \bibfield  {author} {\bibinfo {author} {\bibfnamefont {M.}~\bibnamefont
  {Lyatti}}, \bibinfo {author} {\bibfnamefont {M.~A.}\ \bibnamefont {Wolff}},
  \bibinfo {author} {\bibfnamefont {A.}~\bibnamefont {Savenko}}, \bibinfo
  {author} {\bibfnamefont {M.}~\bibnamefont {Kruth}}, \bibinfo {author}
  {\bibfnamefont {S.}~\bibnamefont {Ferrari}}, \bibinfo {author} {\bibfnamefont
  {U.}~\bibnamefont {Poppe}}, \bibinfo {author} {\bibfnamefont
  {W.}~\bibnamefont {Pernice}}, \bibinfo {author} {\bibfnamefont {R.~E.}\
  \bibnamefont {{Dunin-Borkowski}}}, \ and\ \bibinfo {author} {\bibfnamefont
  {C.}~\bibnamefont {Schuck}},\ }\bibfield  {title} {\enquote {\bibinfo {title}
  {Experimental evidence for hotspot and phase-slip mechanisms of voltage
  switching in ultrathin {YBa$_2$Cu$_3$O$_{7-x}$} nanowires},}\ }\href
  {\doibase 10.1103/PhysRevB.98.054505} {\bibfield  {journal} {\bibinfo
  {journal} {Phys. Rev. B}\ }\textbf {\bibinfo {volume} {98}},\ \bibinfo
  {pages} {054505} (\bibinfo {year} {2018})}\BibitemShut {NoStop}%
\bibitem [{\citenamefont {Rouco}\ \emph {et~al.}(2018)\citenamefont {Rouco},
  \citenamefont {Massarotti}, \citenamefont {Stornaiuolo}, \citenamefont
  {Papari}, \citenamefont {Obradors}, \citenamefont {Puig}, \citenamefont
  {Tafuri},\ and\ \citenamefont {Palau}}]{rouco2018}%
  \BibitemOpen
  \bibfield  {author} {\bibinfo {author} {\bibfnamefont {V{\'i}ctor}\
  \bibnamefont {Rouco}}, \bibinfo {author} {\bibfnamefont {Davide}\
  \bibnamefont {Massarotti}}, \bibinfo {author} {\bibfnamefont {Daniela}\
  \bibnamefont {Stornaiuolo}}, \bibinfo {author} {\bibfnamefont {Gian~Paolo}\
  \bibnamefont {Papari}}, \bibinfo {author} {\bibfnamefont {Xavier}\
  \bibnamefont {Obradors}}, \bibinfo {author} {\bibfnamefont {Teresa}\
  \bibnamefont {Puig}}, \bibinfo {author} {\bibfnamefont {Francesco}\
  \bibnamefont {Tafuri}}, \ and\ \bibinfo {author} {\bibfnamefont {Anna}\
  \bibnamefont {Palau}},\ }\bibfield  {title} {\enquote {\bibinfo {title}
  {Vortex {Lattice Instabilities} in {YBa$_2$Cu$_3$O$_{7-x}$ Nanowires}},}\
  }\href {\doibase 10.3390/ma11020211} {\bibfield  {journal} {\bibinfo
  {journal} {Materials}\ }\textbf {\bibinfo {volume} {11}},\ \bibinfo {pages}
  {211} (\bibinfo {year} {2018})}\BibitemShut {NoStop}%
\bibitem [{\citenamefont {Wu}\ \emph {et~al.}(2015)\citenamefont {Wu},
  \citenamefont {Kuo}, \citenamefont {Jhan}, \citenamefont {Chen},\ and\
  \citenamefont {Jeng}}]{wu2015}%
  \BibitemOpen
  \bibfield  {author} {\bibinfo {author} {\bibfnamefont {C.~H.}\ \bibnamefont
  {Wu}}, \bibinfo {author} {\bibfnamefont {W.~S.}\ \bibnamefont {Kuo}},
  \bibinfo {author} {\bibfnamefont {F.~J.}\ \bibnamefont {Jhan}}, \bibinfo
  {author} {\bibfnamefont {J.~H.}\ \bibnamefont {Chen}}, \ and\ \bibinfo
  {author} {\bibfnamefont {J.~T.}\ \bibnamefont {Jeng}},\ }\bibfield  {title}
  {\enquote {\bibinfo {title} {A {{Nanoscale-Localized Ion Damage Josephson
  Junction Using Focused Ion Beam}} and {{Ion Implanter}}},}\ }\href {\doibase
  10.1166/jnn.2015.9757} {\bibfield  {journal} {\bibinfo  {journal} {J.
  Nanosci. Nanotechnol.}\ }\textbf {\bibinfo {volume} {15}},\ \bibinfo {pages}
  {3728--3732} (\bibinfo {year} {2015})}\BibitemShut {NoStop}%
\bibitem [{\citenamefont {Hilgenkamp}\ and\ \citenamefont
  {Mannhart}(2002)}]{hilgenkamp2002}%
  \BibitemOpen
  \bibfield  {author} {\bibinfo {author} {\bibfnamefont {H.}~\bibnamefont
  {Hilgenkamp}}\ and\ \bibinfo {author} {\bibfnamefont {J.}~\bibnamefont
  {Mannhart}},\ }\bibfield  {title} {\enquote {\bibinfo {title} {Grain
  boundaries in high-{$T_c$} superconductors},}\ }\href {\doibase
  10.1103/RevModPhys.74.485} {\bibfield  {journal} {\bibinfo  {journal} {Rev.
  Mod. Phys.}\ }\textbf {\bibinfo {volume} {74}},\ \bibinfo {pages} {485--549}
  (\bibinfo {year} {2002})}\BibitemShut {NoStop}%
\bibitem [{\citenamefont {Nagel}\ \emph {et~al.}(2011)\citenamefont {Nagel},
  \citenamefont {Konovalenko}, \citenamefont {Kemmler}, \citenamefont {Turad},
  \citenamefont {Werner}, \citenamefont {Kleisz}, \citenamefont {Menzel},
  \citenamefont {Klingeler}, \citenamefont {B{\"u}chner}, \citenamefont
  {Kleiner},\ and\ \citenamefont {Koelle}}]{Nagel11}%
  \BibitemOpen
  \bibfield  {author} {\bibinfo {author} {\bibfnamefont {J.}~\bibnamefont
  {Nagel}}, \bibinfo {author} {\bibfnamefont {K.~B.}\ \bibnamefont
  {Konovalenko}}, \bibinfo {author} {\bibfnamefont {M.}~\bibnamefont
  {Kemmler}}, \bibinfo {author} {\bibfnamefont {M.}~\bibnamefont {Turad}},
  \bibinfo {author} {\bibfnamefont {R.}~\bibnamefont {Werner}}, \bibinfo
  {author} {\bibfnamefont {E.}~\bibnamefont {Kleisz}}, \bibinfo {author}
  {\bibfnamefont {S.}~\bibnamefont {Menzel}}, \bibinfo {author} {\bibfnamefont
  {R.}~\bibnamefont {Klingeler}}, \bibinfo {author} {\bibfnamefont
  {B.}~\bibnamefont {B{\"u}chner}}, \bibinfo {author} {\bibfnamefont
  {R.}~\bibnamefont {Kleiner}}, \ and\ \bibinfo {author} {\bibfnamefont
  {D.}~\bibnamefont {Koelle}},\ }\bibfield  {title} {\enquote {\bibinfo {title}
  {Resistively shunted {YBa$_2$Cu$_3$O$_7$} grain boundary junctions and
  low-noise {SQUIDs} patterned by a focused ion beam down to 80\,nm
  linewidth},}\ }\href {\doibase 10.1088/0953-2048/24/1/015015} {\bibfield
  {journal} {\bibinfo  {journal} {Supercond. Sci. Technol.}\ }\textbf {\bibinfo
  {volume} {24}},\ \bibinfo {pages} {015015} (\bibinfo {year}
  {2011})}\BibitemShut {NoStop}%
\bibitem [{\citenamefont {Cybart}\ \emph {et~al.}(2016)\citenamefont {Cybart},
  \citenamefont {Bali}, \citenamefont {Hlawacek}, \citenamefont {R\"{o}der},\
  and\ \citenamefont {Fassbender}}]{Cybart17}%
  \BibitemOpen
  \bibfield  {author} {\bibinfo {author} {\bibfnamefont {S.}~\bibnamefont
  {Cybart}}, \bibinfo {author} {\bibfnamefont {R.}~\bibnamefont {Bali}},
  \bibinfo {author} {\bibfnamefont {G.}~\bibnamefont {Hlawacek}}, \bibinfo
  {author} {\bibfnamefont {F.}~\bibnamefont {R\"{o}der}}, \ and\ \bibinfo
  {author} {\bibfnamefont {J.}~\bibnamefont {Fassbender}},\ }\enquote {\bibinfo
  {title} {Focused {H}elium and {N}eon {I}on {B}eam {M}odification of
  {H}igh-{$T_c$} {S}uperconductors and {M}agnetic {M}aterials},}\ in\
  \href@noop {} {\emph {\bibinfo {booktitle} {Helium Ion Microscopy}}},\
  \bibinfo {editor} {edited by\ \bibinfo {editor} {\bibfnamefont
  {G.}~\bibnamefont {Hlawacek}}\ and\ \bibinfo {editor} {\bibfnamefont
  {A.}~\bibnamefont {G\"{o}lzh\"{a}user}}}\ (\bibinfo  {publisher} {Springer
  International Publishing},\ \bibinfo {address} {Cham (Switzerland)},\
  \bibinfo {year} {2016})\ Chap.~\bibinfo {chapter} {17}, pp.\ \bibinfo {pages}
  {415--455}\BibitemShut {NoStop}%
\bibitem [{\citenamefont {Lang}\ \emph {et~al.}(2006)\citenamefont {Lang},
  \citenamefont {Dineva}, \citenamefont {Marksteiner}, \citenamefont
  {Enzenhofer}, \citenamefont {Siraj}, \citenamefont {Peruzzi}, \citenamefont
  {Pedarnig}, \citenamefont {B{\"a}uerle}, \citenamefont {Korntner},
  \citenamefont {Cekan}, \citenamefont {Platzgummer},\ and\ \citenamefont
  {Loeschner}}]{lang2006}%
  \BibitemOpen
  \bibfield  {author} {\bibinfo {author} {\bibfnamefont {W.}~\bibnamefont
  {Lang}}, \bibinfo {author} {\bibfnamefont {M.}~\bibnamefont {Dineva}},
  \bibinfo {author} {\bibfnamefont {M.}~\bibnamefont {Marksteiner}}, \bibinfo
  {author} {\bibfnamefont {T.}~\bibnamefont {Enzenhofer}}, \bibinfo {author}
  {\bibfnamefont {K.}~\bibnamefont {Siraj}}, \bibinfo {author} {\bibfnamefont
  {M.}~\bibnamefont {Peruzzi}}, \bibinfo {author} {\bibfnamefont {J.~D.}\
  \bibnamefont {Pedarnig}}, \bibinfo {author} {\bibfnamefont {D.}~\bibnamefont
  {B{\"a}uerle}}, \bibinfo {author} {\bibfnamefont {R.}~\bibnamefont
  {Korntner}}, \bibinfo {author} {\bibfnamefont {E.}~\bibnamefont {Cekan}},
  \bibinfo {author} {\bibfnamefont {E.}~\bibnamefont {Platzgummer}}, \ and\
  \bibinfo {author} {\bibfnamefont {H.}~\bibnamefont {Loeschner}},\ }\bibfield
  {title} {\enquote {\bibinfo {title} {Ion-beam direct-structuring of
  high-temperature superconductors},}\ }\href {\doibase
  10.1016/j.mee.2006.01.091} {\bibfield  {journal} {\bibinfo  {journal}
  {Microelectron. Eng.}\ }\textbf {\bibinfo {volume} {83}},\ \bibinfo {pages}
  {1495--1498} (\bibinfo {year} {2006})}\BibitemShut {NoStop}%
\bibitem [{\citenamefont {Zaluzhnyy}\ \emph {et~al.}(2024)\citenamefont
  {Zaluzhnyy}, \citenamefont {Goteti}, \citenamefont {Stoychev}, \citenamefont
  {Basak}, \citenamefont {Lamb}, \citenamefont {Kisiel}, \citenamefont {Zhou},
  \citenamefont {Cai}, \citenamefont {Holt}, \citenamefont {Beeman},
  \citenamefont {Cho}, \citenamefont {Cybart}, \citenamefont {Shpyrko},
  \citenamefont {Dynes},\ and\ \citenamefont {Frano}}]{zaluzhnyy2024}%
  \BibitemOpen
  \bibfield  {author} {\bibinfo {author} {\bibfnamefont {Ivan~A.}\ \bibnamefont
  {Zaluzhnyy}}, \bibinfo {author} {\bibfnamefont {Uday}\ \bibnamefont
  {Goteti}}, \bibinfo {author} {\bibfnamefont {Boyan~K.}\ \bibnamefont
  {Stoychev}}, \bibinfo {author} {\bibfnamefont {Rourav}\ \bibnamefont
  {Basak}}, \bibinfo {author} {\bibfnamefont {Erik~S.}\ \bibnamefont {Lamb}},
  \bibinfo {author} {\bibfnamefont {Elliot}\ \bibnamefont {Kisiel}}, \bibinfo
  {author} {\bibfnamefont {Tao}\ \bibnamefont {Zhou}}, \bibinfo {author}
  {\bibfnamefont {Zhonghou}\ \bibnamefont {Cai}}, \bibinfo {author}
  {\bibfnamefont {Martin~V.}\ \bibnamefont {Holt}}, \bibinfo {author}
  {\bibfnamefont {Jeffrey~W.}\ \bibnamefont {Beeman}}, \bibinfo {author}
  {\bibfnamefont {Ethan~Y.}\ \bibnamefont {Cho}}, \bibinfo {author}
  {\bibfnamefont {Shane}\ \bibnamefont {Cybart}}, \bibinfo {author}
  {\bibfnamefont {Oleg~G.}\ \bibnamefont {Shpyrko}}, \bibinfo {author}
  {\bibfnamefont {Robert}\ \bibnamefont {Dynes}}, \ and\ \bibinfo {author}
  {\bibfnamefont {Alex}\ \bibnamefont {Frano}},\ }\bibfield  {title} {\enquote
  {\bibinfo {title} {Structural changes in {YBa$_2$Cu$_3$O$_7$} thin films
  modified with {He}$^+$-focused ion beam for high-temperature superconductive
  nanoelectronics},}\ }\href {\doibase 10.1021/acsanm.4c00247} {\bibfield
  {journal} {\bibinfo  {journal} {ACS Appl. Nano Mater.}\ }\textbf {\bibinfo
  {volume} {7}},\ \bibinfo {pages} {15943--15949} (\bibinfo {year}
  {2024})}\BibitemShut {NoStop}%
\bibitem [{\citenamefont {Cybart}\ \emph {et~al.}(2015)\citenamefont {Cybart},
  \citenamefont {Cho}, \citenamefont {Wong}, \citenamefont {Wehlin},
  \citenamefont {Ma}, \citenamefont {Huynh},\ and\ \citenamefont
  {Dynes}}]{cybart2015}%
  \BibitemOpen
  \bibfield  {author} {\bibinfo {author} {\bibfnamefont {Shane~A.}\
  \bibnamefont {Cybart}}, \bibinfo {author} {\bibfnamefont {E.~Y.}\
  \bibnamefont {Cho}}, \bibinfo {author} {\bibfnamefont {T.~J.}\ \bibnamefont
  {Wong}}, \bibinfo {author} {\bibfnamefont {Bj{\"o}rn~H.}\ \bibnamefont
  {Wehlin}}, \bibinfo {author} {\bibfnamefont {Meng~K.}\ \bibnamefont {Ma}},
  \bibinfo {author} {\bibfnamefont {Chuong}\ \bibnamefont {Huynh}}, \ and\
  \bibinfo {author} {\bibfnamefont {R.~C.}\ \bibnamefont {Dynes}},\ }\bibfield
  {title} {\enquote {\bibinfo {title} {Nano {J}osephson superconducting tunnel
  junctions in {YBa$_2$Cu$_3$O$_{7-\delta}$} directly patterned with a focused
  helium ion beam},}\ }\href {\doibase 10.1038/nnano.2015.76} {\bibfield
  {journal} {\bibinfo  {journal} {Nat. Nanotechnol.}\ }\textbf {\bibinfo
  {volume} {10}},\ \bibinfo {pages} {598--602} (\bibinfo {year}
  {2015})}\BibitemShut {NoStop}%
\bibitem [{\citenamefont {M\"{u}ller}\ \emph {et~al.}(2019)\citenamefont
  {M\"{u}ller}, \citenamefont {Karrer}, \citenamefont {Limberger},
  \citenamefont {Becker}, \citenamefont {Schr\"{o}ppel}, \citenamefont
  {Burkhardt}, \citenamefont {Kleiner}, \citenamefont {Goldobin},\ and\
  \citenamefont {Koelle}}]{mueller19}%
  \BibitemOpen
  \bibfield  {author} {\bibinfo {author} {\bibfnamefont {B.}~\bibnamefont
  {M\"{u}ller}}, \bibinfo {author} {\bibfnamefont {M.}~\bibnamefont {Karrer}},
  \bibinfo {author} {\bibfnamefont {F.}~\bibnamefont {Limberger}}, \bibinfo
  {author} {\bibfnamefont {M.}~\bibnamefont {Becker}}, \bibinfo {author}
  {\bibfnamefont {B.}~\bibnamefont {Schr\"{o}ppel}}, \bibinfo {author}
  {\bibfnamefont {C.~J.}\ \bibnamefont {Burkhardt}}, \bibinfo {author}
  {\bibfnamefont {R.}~\bibnamefont {Kleiner}}, \bibinfo {author} {\bibfnamefont
  {E.}~\bibnamefont {Goldobin}}, \ and\ \bibinfo {author} {\bibfnamefont
  {D.}~\bibnamefont {Koelle}},\ }\bibfield  {title} {\enquote {\bibinfo {title}
  {Josephson junctions and {SQUIDs} created by focused helium-ion-beam
  irradiation of {YBa$_2$Cu$_3$O$_7$}},}\ }\href {\doibase
  10.1103/PhysRevApplied.11.044082} {\bibfield  {journal} {\bibinfo  {journal}
  {Phys. Rev. Appl.}\ }\textbf {\bibinfo {volume} {11}},\ \bibinfo {pages}
  {044082} (\bibinfo {year} {2019})}\BibitemShut {NoStop}%
\bibitem [{\citenamefont {Karrer}\ \emph {et~al.}(2024)\citenamefont {Karrer},
  \citenamefont {Wurster}, \citenamefont {Linek}, \citenamefont {Meichsner},
  \citenamefont {Kleiner}, \citenamefont {Goldobin},\ and\ \citenamefont
  {Koelle}}]{Karrer24}%
  \BibitemOpen
  \bibfield  {author} {\bibinfo {author} {\bibfnamefont {M.}~\bibnamefont
  {Karrer}}, \bibinfo {author} {\bibfnamefont {K.}~\bibnamefont {Wurster}},
  \bibinfo {author} {\bibfnamefont {J.}~\bibnamefont {Linek}}, \bibinfo
  {author} {\bibfnamefont {M.}~\bibnamefont {Meichsner}}, \bibinfo {author}
  {\bibfnamefont {R.}~\bibnamefont {Kleiner}}, \bibinfo {author} {\bibfnamefont
  {E.}~\bibnamefont {Goldobin}}, \ and\ \bibinfo {author} {\bibfnamefont
  {D.}~\bibnamefont {Koelle}},\ }\bibfield  {title} {\enquote {\bibinfo {title}
  {Temporal evolution of electric transport properties of
  {YBa$_2$Cu$_3$O$_{7-\delta}$} {J}osephson junctions produced by
  focused-helium-ion-beam irradiation},}\ }\href {\doibase
  10.1103/PhysRevApplied.21.014065} {\bibfield  {journal} {\bibinfo  {journal}
  {Phys. Rev. Appl.}\ }\textbf {\bibinfo {volume} {21}},\ \bibinfo {pages}
  {014065} (\bibinfo {year} {2024})}\BibitemShut {NoStop}%
\bibitem [{\citenamefont {Pr{\"o}pper}\ \emph {et~al.}(2024)\citenamefont
  {Pr{\"o}pper}, \citenamefont {Hanisch}, \citenamefont {Schmid}, \citenamefont
  {Ritter}, \citenamefont {Neumann}, \citenamefont {Goldobin}, \citenamefont
  {Koelle}, \citenamefont {Kleiner}, \citenamefont {Schilling},\ and\
  \citenamefont {Hampel}}]{propper2024}%
  \BibitemOpen
  \bibfield  {author} {\bibinfo {author} {\bibfnamefont {M.}~\bibnamefont
  {Pr{\"o}pper}}, \bibinfo {author} {\bibfnamefont {D.}~\bibnamefont
  {Hanisch}}, \bibinfo {author} {\bibfnamefont {C.}~\bibnamefont {Schmid}},
  \bibinfo {author} {\bibfnamefont {P.~J.}\ \bibnamefont {Ritter}}, \bibinfo
  {author} {\bibfnamefont {M.}~\bibnamefont {Neumann}}, \bibinfo {author}
  {\bibfnamefont {E.}~\bibnamefont {Goldobin}}, \bibinfo {author}
  {\bibfnamefont {D.}~\bibnamefont {Koelle}}, \bibinfo {author} {\bibfnamefont
  {R.}~\bibnamefont {Kleiner}}, \bibinfo {author} {\bibfnamefont
  {M.}~\bibnamefont {Schilling}}, \ and\ \bibinfo {author} {\bibfnamefont
  {B.}~\bibnamefont {Hampel}},\ }\bibfield  {title} {\enquote {\bibinfo {title}
  {{THz} properties of {He-FIB} {YBa$_2$Cu$_3$O$_{7-x}$} {J}osephson
  junctions},}\ }\href {\doibase 10.1109/TASC.2024.3353143} {\bibfield
  {journal} {\bibinfo  {journal} {IEEE Trans. Appl. Supercond.}\ }\textbf
  {\bibinfo {volume} {34}},\ \bibinfo {pages} {1--5} (\bibinfo {year}
  {2024})}\BibitemShut {NoStop}%
\bibitem [{\citenamefont {Cho}\ \emph {et~al.}(2018{\natexlab{a}})\citenamefont
  {Cho}, \citenamefont {Zhou}, \citenamefont {Cho},\ and\ \citenamefont
  {Cybart}}]{cho2018}%
  \BibitemOpen
  \bibfield  {author} {\bibinfo {author} {\bibfnamefont {Ethan~Y.}\
  \bibnamefont {Cho}}, \bibinfo {author} {\bibfnamefont {Yuchao~W.}\
  \bibnamefont {Zhou}}, \bibinfo {author} {\bibfnamefont {Jennifer~Y.}\
  \bibnamefont {Cho}}, \ and\ \bibinfo {author} {\bibfnamefont {Shane~A.}\
  \bibnamefont {Cybart}},\ }\bibfield  {title} {\enquote {\bibinfo {title}
  {Superconducting nano {{Josephson}} junctions patterned with a focused helium
  ion beam},}\ }\href {\doibase 10.1063/1.5042105} {\bibfield  {journal}
  {\bibinfo  {journal} {Appl. Phys. Lett.}\ }\textbf {\bibinfo {volume}
  {113}},\ \bibinfo {pages} {022604} (\bibinfo {year}
  {2018}{\natexlab{a}})}\BibitemShut {NoStop}%
\bibitem [{\citenamefont {Cho}\ \emph {et~al.}(2018{\natexlab{b}})\citenamefont
  {Cho}, \citenamefont {Li}, \citenamefont {LeFebvre}, \citenamefont {Zhou},
  \citenamefont {Dynes},\ and\ \citenamefont {Cybart}}]{cho2018a}%
  \BibitemOpen
  \bibfield  {author} {\bibinfo {author} {\bibfnamefont {Ethan~Y.}\
  \bibnamefont {Cho}}, \bibinfo {author} {\bibfnamefont {Hao}\ \bibnamefont
  {Li}}, \bibinfo {author} {\bibfnamefont {Jay~C.}\ \bibnamefont {LeFebvre}},
  \bibinfo {author} {\bibfnamefont {Yuchao~W.}\ \bibnamefont {Zhou}}, \bibinfo
  {author} {\bibfnamefont {R.~C.}\ \bibnamefont {Dynes}}, \ and\ \bibinfo
  {author} {\bibfnamefont {Shane~A.}\ \bibnamefont {Cybart}},\ }\bibfield
  {title} {\enquote {\bibinfo {title} {Direct-coupled micro-magnetometer with
  {{Y-Ba-Cu-O}} nano-slit {{SQUID}} fabricated with a focused helium ion
  beam},}\ }\href {\doibase 10.1063/1.5048776} {\bibfield  {journal} {\bibinfo
  {journal} {Appl. Phys. Lett.}\ }\textbf {\bibinfo {volume} {113}},\ \bibinfo
  {pages} {162602} (\bibinfo {year} {2018}{\natexlab{b}})}\BibitemShut
  {NoStop}%
\bibitem [{\citenamefont {Li}\ \emph {et~al.}(2019)\citenamefont {Li},
  \citenamefont {Cho}, \citenamefont {Cai}, \citenamefont {Wang}, \citenamefont
  {McCoy},\ and\ \citenamefont {Cybart}}]{li2019}%
  \BibitemOpen
  \bibfield  {author} {\bibinfo {author} {\bibfnamefont {Hao}\ \bibnamefont
  {Li}}, \bibinfo {author} {\bibfnamefont {Ethan~Y.}\ \bibnamefont {Cho}},
  \bibinfo {author} {\bibfnamefont {Han}\ \bibnamefont {Cai}}, \bibinfo
  {author} {\bibfnamefont {Yan-Ting}\ \bibnamefont {Wang}}, \bibinfo {author}
  {\bibfnamefont {Stephen~J.}\ \bibnamefont {McCoy}}, \ and\ \bibinfo {author}
  {\bibfnamefont {Shane~A.}\ \bibnamefont {Cybart}},\ }\bibfield  {title}
  {\enquote {\bibinfo {title} {Inductance investigation of
  {YBa$_2$Cu$_3$O$_{7-\delta}$} nano-slit {SQUIDs} fabricated with a focused
  helium ion beam},}\ }\href {\doibase 10.1109/TASC.2019.2898692} {\bibfield
  {journal} {\bibinfo  {journal} {IEEE Trans. Appl. Supercond.}\ }\textbf
  {\bibinfo {volume} {29}},\ \bibinfo {pages} {1--4} (\bibinfo {year}
  {2019})}\BibitemShut {NoStop}%
\bibitem [{\citenamefont {Schmid}\ \emph {et~al.}(2025)\citenamefont {Schmid},
  \citenamefont {Jozani}, \citenamefont {Kleiner}, \citenamefont {Koelle},\
  and\ \citenamefont {Goldobin}}]{Schmid:2025:He-FIB:YBCO-JJD}%
  \BibitemOpen
  \bibfield  {author} {\bibinfo {author} {\bibfnamefont {Christoph}\
  \bibnamefont {Schmid}}, \bibinfo {author} {\bibfnamefont {Alireza}\
  \bibnamefont {Jozani}}, \bibinfo {author} {\bibfnamefont {Reinhold}\
  \bibnamefont {Kleiner}}, \bibinfo {author} {\bibfnamefont {Dieter}\
  \bibnamefont {Koelle}}, \ and\ \bibinfo {author} {\bibfnamefont {Edward}\
  \bibnamefont {Goldobin}},\ }\bibfield  {title} {\enquote {\bibinfo {title}
  {{\YBCO} {J}osephson diode fabricated by focused-helium-ion-beam
  irradiation},}\ }\href {\doibase 10.1103/vqhx-16ss} {\bibfield  {journal}
  {\bibinfo  {journal} {Phys. Rev. Appl.}\ }\textbf {\bibinfo {volume} {24}},\
  \bibinfo {pages} {014041} (\bibinfo {year} {2025})}\BibitemShut {NoStop}%
\bibitem [{\citenamefont {Schilling}\ \emph {et~al.}(2013)\citenamefont
  {Schilling}, \citenamefont {Guillaume}, \citenamefont {Scholtyssek},\ and\
  \citenamefont {Ludwig}}]{schilling2013}%
  \BibitemOpen
  \bibfield  {author} {\bibinfo {author} {\bibfnamefont {M.}~\bibnamefont
  {Schilling}}, \bibinfo {author} {\bibfnamefont {A.}~\bibnamefont
  {Guillaume}}, \bibinfo {author} {\bibfnamefont {J.~M.}\ \bibnamefont
  {Scholtyssek}}, \ and\ \bibinfo {author} {\bibfnamefont {F.}~\bibnamefont
  {Ludwig}},\ }\bibfield  {title} {\enquote {\bibinfo {title} {Design of
  experiments for highly reproducible pulsed laser deposition of
  {YBa$_2$Cu$_3$O$_{7-\delta}$}},}\ }\href {\doibase
  10.1088/0022-3727/47/3/034008} {\bibfield  {journal} {\bibinfo  {journal} {J.
  Phys. D: Appl. Phys.}\ }\textbf {\bibinfo {volume} {47}},\ \bibinfo {pages}
  {034008} (\bibinfo {year} {2013})}\BibitemShut {NoStop}%
\bibitem [{\citenamefont {Sarkar}\ and\ \citenamefont
  {Cybart}(2023)}]{sarkar2023}%
  \BibitemOpen
  \bibfield  {author} {\bibinfo {author} {\bibfnamefont {Nirjhar}\ \bibnamefont
  {Sarkar}}\ and\ \bibinfo {author} {\bibfnamefont {Shane~A.}\ \bibnamefont
  {Cybart}},\ }\bibfield  {title} {\enquote {\bibinfo {title} {Design {{Rules}}
  for {{Resistors}}, {{Capacitors}} and {{Inductors Fabricated From Single
  Layer Y-Ba-Cu-O Thin Films With Focused Helium Ion Beam Irradiation}}},}\
  }\href {\doibase 10.1109/TASC.2023.3257119} {\bibfield  {journal} {\bibinfo
  {journal} {IEEE Trans. Appl. Supercond.}\ }\textbf {\bibinfo {volume} {33}},\
  \bibinfo {pages} {1--5} (\bibinfo {year} {2023})}\BibitemShut {NoStop}%
\bibitem [{\citenamefont {Lesueur}\ \emph {et~al.}(1993)\citenamefont
  {Lesueur}, \citenamefont {Dumoulin}, \citenamefont {Quillet},\ and\
  \citenamefont {Radcliffe}}]{Lesueur:1993:YBCO:SIT(He-ions)}%
  \BibitemOpen
  \bibfield  {author} {\bibinfo {author} {\bibfnamefont {J.}~\bibnamefont
  {Lesueur}}, \bibinfo {author} {\bibfnamefont {L.}~\bibnamefont {Dumoulin}},
  \bibinfo {author} {\bibfnamefont {S.}~\bibnamefont {Quillet}}, \ and\
  \bibinfo {author} {\bibfnamefont {J.}~\bibnamefont {Radcliffe}},\ }\bibfield
  {title} {\enquote {\bibinfo {title} {Ion-beam induced metal insulator
  transition in {YBCO} films},}\ }\href {\doibase
  https://doi.org/10.1016/0925-8388(93)90792-L} {\bibfield  {journal} {\bibinfo
   {journal} {J. Alloys Compd.}\ }\textbf {\bibinfo {volume} {195}},\ \bibinfo
  {pages} {527--530} (\bibinfo {year} {1993})}\BibitemShut {NoStop}%
\bibitem [{\citenamefont {Pr{\"o}pper}\ \emph {et~al.}(2025)\citenamefont
  {Pr{\"o}pper}, \citenamefont {Hanisch}, \citenamefont {Schmid}, \citenamefont
  {Neumann}, \citenamefont {Ritter}, \citenamefont {Tucholke}, \citenamefont
  {Goldobin}, \citenamefont {Koelle}, \citenamefont {Kleiner}, \citenamefont
  {Schilling},\ and\ \citenamefont {Hampel}}]{propper2025}%
  \BibitemOpen
  \bibfield  {author} {\bibinfo {author} {\bibfnamefont {Max}\ \bibnamefont
  {Pr{\"o}pper}}, \bibinfo {author} {\bibfnamefont {Dominik}\ \bibnamefont
  {Hanisch}}, \bibinfo {author} {\bibfnamefont {Christoph}\ \bibnamefont
  {Schmid}}, \bibinfo {author} {\bibfnamefont {Marius}\ \bibnamefont
  {Neumann}}, \bibinfo {author} {\bibfnamefont {Paul~Julius}\ \bibnamefont
  {Ritter}}, \bibinfo {author} {\bibfnamefont {Marc-Andr{\'e}}\ \bibnamefont
  {Tucholke}}, \bibinfo {author} {\bibfnamefont {Edward}\ \bibnamefont
  {Goldobin}}, \bibinfo {author} {\bibfnamefont {Dieter}\ \bibnamefont
  {Koelle}}, \bibinfo {author} {\bibfnamefont {Reinhold}\ \bibnamefont
  {Kleiner}}, \bibinfo {author} {\bibfnamefont {Meinhard}\ \bibnamefont
  {Schilling}}, \ and\ \bibinfo {author} {\bibfnamefont {Benedikt}\
  \bibnamefont {Hampel}},\ }\bibfield  {title} {\enquote {\bibinfo {title}
  {High-{$T_c$} {J}osephson junction arrays fabricated by {He-FIB}},}\ }\href
  {\doibase 10.1109/TASC.2024.3516741} {\bibfield  {journal} {\bibinfo
  {journal} {IEEE Trans. Appl. Supercond.}\ }\textbf {\bibinfo {volume} {35}},\
  \bibinfo {pages} {1--5} (\bibinfo {year} {2025})}\BibitemShut {NoStop}%
\bibitem [{\citenamefont {Bardeen}(1962)}]{bardeen1962}%
  \BibitemOpen
  \bibfield  {author} {\bibinfo {author} {\bibfnamefont {John}\ \bibnamefont
  {Bardeen}},\ }\bibfield  {title} {\enquote {\bibinfo {title} {Critical
  {{Fields}} and {{Currents}} in {{Superconductors}}},}\ }\href {\doibase
  10.1103/RevModPhys.34.667} {\bibfield  {journal} {\bibinfo  {journal} {Rev.
  Mod. Phys.}\ }\textbf {\bibinfo {volume} {34}},\ \bibinfo {pages} {667--681}
  (\bibinfo {year} {1962})}\BibitemShut {NoStop}%
\bibitem [{Note1()}]{Note1}%
  \BibitemOpen
  \bibinfo {note} {Effective area $A_\protect \mathrm {eff}$ is calculated as
  fluxoid $\protect \mathaccentV {tilde}07E{\Phi }$ induced around the SQUID
  hole (output of the simulation) divided by the applied field $B$ (input
  parameter), \protect \textit {i.e.}\protect \xspace , $A_\protect \mathrm
  {eff}=\protect \mathaccentV {tilde}07E{\Phi }/B$.}\BibitemShut {Stop}%
\bibitem [{\citenamefont {Khapaev}\ \emph {et~al.}(2001)\citenamefont
  {Khapaev}, \citenamefont {{Kidiyarova-Shevchenko}}, \citenamefont
  {Magnelind},\ and\ \citenamefont {Kupriyanov}}]{khapaev2001}%
  \BibitemOpen
  \bibfield  {author} {\bibinfo {author} {\bibfnamefont {M.M.}\ \bibnamefont
  {Khapaev}}, \bibinfo {author} {\bibfnamefont {{\relax A.Yu}.}~\bibnamefont
  {{Kidiyarova-Shevchenko}}}, \bibinfo {author} {\bibfnamefont
  {P.}~\bibnamefont {Magnelind}}, \ and\ \bibinfo {author} {\bibfnamefont
  {{\relax M.Yu}.}~\bibnamefont {Kupriyanov}},\ }\bibfield  {title} {\enquote
  {\bibinfo {title} {{{3D-MLSI}}: Software package for inductance calculation
  in multilayer superconducting integrated circuits},}\ }\href {\doibase
  10.1109/77.919537} {\bibfield  {journal} {\bibinfo  {journal} {IEEE Trans.
  Appl. Supercond.}\ }\textbf {\bibinfo {volume} {11}},\ \bibinfo {pages}
  {1090--1093} (\bibinfo {year} {2001})}\BibitemShut {NoStop}%
\bibitem [{\citenamefont {Tesche}\ and\ \citenamefont
  {Clarke}(1977)}]{tesche1977}%
  \BibitemOpen
  \bibfield  {author} {\bibinfo {author} {\bibfnamefont {Claudia~D.}\
  \bibnamefont {Tesche}}\ and\ \bibinfo {author} {\bibfnamefont {John}\
  \bibnamefont {Clarke}},\ }\bibfield  {title} {\enquote {\bibinfo {title} {Dc
  {{SQUID}}: {{Noise}} and optimization},}\ }\href {\doibase
  10.1007/BF00655097} {\bibfield  {journal} {\bibinfo  {journal} {J. Low Temp.
  Phys.}\ }\textbf {\bibinfo {volume} {29}},\ \bibinfo {pages} {301--331}
  (\bibinfo {year} {1977})}\BibitemShut {NoStop}%
\bibitem [{\citenamefont {Behr}\ \emph {et~al.}(2012)\citenamefont {Behr},
  \citenamefont {Kieler}, \citenamefont {Kohlmann}, \citenamefont
  {M{\"u}ller},\ and\ \citenamefont {Palafox}}]{behr2012}%
  \BibitemOpen
  \bibfield  {author} {\bibinfo {author} {\bibfnamefont {Ralf}\ \bibnamefont
  {Behr}}, \bibinfo {author} {\bibfnamefont {Oliver}\ \bibnamefont {Kieler}},
  \bibinfo {author} {\bibfnamefont {Johannes}\ \bibnamefont {Kohlmann}},
  \bibinfo {author} {\bibfnamefont {Franz}\ \bibnamefont {M{\"u}ller}}, \ and\
  \bibinfo {author} {\bibfnamefont {Luis}\ \bibnamefont {Palafox}},\ }\bibfield
   {title} {\enquote {\bibinfo {title} {Development and metrological
  applications of {{Josephson}} arrays at {{PTB}}},}\ }\href {\doibase
  10.1088/0957-0233/23/12/124002} {\bibfield  {journal} {\bibinfo  {journal}
  {Meas. Sci. Technol.}\ }\textbf {\bibinfo {volume} {23}},\ \bibinfo {pages}
  {124002} (\bibinfo {year} {2012})}\BibitemShut {NoStop}%
\bibitem [{\citenamefont {R{\"u}fenacht}\ \emph {et~al.}(2018)\citenamefont
  {R{\"u}fenacht}, \citenamefont {{Flowers-Jacobs}},\ and\ \citenamefont
  {Benz}}]{rufenacht2018a}%
  \BibitemOpen
  \bibfield  {author} {\bibinfo {author} {\bibfnamefont {Alain}\ \bibnamefont
  {R{\"u}fenacht}}, \bibinfo {author} {\bibfnamefont {Nathan~E}\ \bibnamefont
  {{Flowers-Jacobs}}}, \ and\ \bibinfo {author} {\bibfnamefont {Samuel~P}\
  \bibnamefont {Benz}},\ }\bibfield  {title} {\enquote {\bibinfo {title}
  {Impact of the latest generation of {{Josephson}} voltage standards in ac and
  dc electric metrology},}\ }\href {\doibase 10.1088/1681-7575/aad41a}
  {\bibfield  {journal} {\bibinfo  {journal} {Metrologia}\ }\textbf {\bibinfo
  {volume} {55}},\ \bibinfo {pages} {S152--S173} (\bibinfo {year}
  {2018})}\BibitemShut {NoStop}%
\bibitem [{\citenamefont {Tollk{\"u}hn}\ \emph {et~al.}(2022)\citenamefont
  {Tollk{\"u}hn}, \citenamefont {Ritter}, \citenamefont {Schilling},\ and\
  \citenamefont {Hampel}}]{tollkuhn2022}%
  \BibitemOpen
  \bibfield  {author} {\bibinfo {author} {\bibfnamefont {M.}~\bibnamefont
  {Tollk{\"u}hn}}, \bibinfo {author} {\bibfnamefont {P.~J.}\ \bibnamefont
  {Ritter}}, \bibinfo {author} {\bibfnamefont {M.}~\bibnamefont {Schilling}}, \
  and\ \bibinfo {author} {\bibfnamefont {B.}~\bibnamefont {Hampel}},\
  }\bibfield  {title} {\enquote {\bibinfo {title} {{{THz}} microscope for
  three-dimensional imaging with superconducting {{Josephson}} junctions},}\
  }\href {\doibase 10.1063/5.0084207} {\bibfield  {journal} {\bibinfo
  {journal} {Rev. Sci. Instrum.}\ }\textbf {\bibinfo {volume} {93}},\ \bibinfo
  {pages} {043708} (\bibinfo {year} {2022})}\BibitemShut {NoStop}%
\bibitem [{\citenamefont {Ritter}\ \emph {et~al.}(2022)\citenamefont {Ritter},
  \citenamefont {Tollk{\"u}hn}, \citenamefont {Schilling},\ and\ \citenamefont
  {Hampel}}]{ritter2022}%
  \BibitemOpen
  \bibfield  {author} {\bibinfo {author} {\bibfnamefont {P.J.}\ \bibnamefont
  {Ritter}}, \bibinfo {author} {\bibfnamefont {M.}~\bibnamefont
  {Tollk{\"u}hn}}, \bibinfo {author} {\bibfnamefont {M.}~\bibnamefont
  {Schilling}}, \ and\ \bibinfo {author} {\bibfnamefont {B.}~\bibnamefont
  {Hampel}},\ }\bibfield  {title} {\enquote {\bibinfo {title} {Characterization
  of {{Multi-Frequency Emission}} of {{Far-Infrared Laser}} with {{Josephson
  Junctions}} in a {{THz Microscope}}},}\ }in\ \href {\doibase
  10.1109/IRMMW-THz50927.2022.9895908} {\emph {\bibinfo {booktitle}
  {IRMMW-THz}}}\ (\bibinfo {year} {2022})\ pp.\ \bibinfo {pages}
  {1--2}\BibitemShut {NoStop}%
\bibitem [{\citenamefont {De~Simoni}\ \emph {et~al.}(2020)\citenamefont
  {De~Simoni}, \citenamefont {Puglia},\ and\ \citenamefont
  {Giazotto}}]{desimoni2020}%
  \BibitemOpen
  \bibfield  {author} {\bibinfo {author} {\bibfnamefont {G.}~\bibnamefont
  {De~Simoni}}, \bibinfo {author} {\bibfnamefont {C.}~\bibnamefont {Puglia}}, \
  and\ \bibinfo {author} {\bibfnamefont {F.}~\bibnamefont {Giazotto}},\
  }\bibfield  {title} {\enquote {\bibinfo {title} {Niobium {{Dayem}}
  nano-bridge {{Josephson}} gate-controlled transistors},}\ }\href {\doibase
  10.1063/5.0011304} {\bibfield  {journal} {\bibinfo  {journal} {Appl. Phys.
  Lett.}\ }\textbf {\bibinfo {volume} {116}},\ \bibinfo {pages} {242601}
  (\bibinfo {year} {2020})}\BibitemShut {NoStop}%
\bibitem [{\citenamefont {Lang}\ \emph {et~al.}(2005)\citenamefont {Lang},
  \citenamefont {Puica}, \citenamefont {Peruzzi}, \citenamefont {Lemmermann},
  \citenamefont {Pedarnig},\ and\ \citenamefont {B{\"a}uerle}}]{lang2005}%
  \BibitemOpen
  \bibfield  {author} {\bibinfo {author} {\bibfnamefont {W.}~\bibnamefont
  {Lang}}, \bibinfo {author} {\bibfnamefont {I.}~\bibnamefont {Puica}},
  \bibinfo {author} {\bibfnamefont {M.}~\bibnamefont {Peruzzi}}, \bibinfo
  {author} {\bibfnamefont {K.}~\bibnamefont {Lemmermann}}, \bibinfo {author}
  {\bibfnamefont {J.~D.}\ \bibnamefont {Pedarnig}}, \ and\ \bibinfo {author}
  {\bibfnamefont {D.}~\bibnamefont {B{\"a}uerle}},\ }\bibfield  {title}
  {\enquote {\bibinfo {title} {Depairing current and superconducting transition
  of {{YBCO}} at intense pulsed currents},}\ }\href {\doibase
  10.1002/pssc.200460801} {\bibfield  {journal} {\bibinfo  {journal} {Phys.
  Status Solidi C}\ }\textbf {\bibinfo {volume} {2}},\ \bibinfo {pages}
  {1615--1624} (\bibinfo {year} {2005})}\BibitemShut {NoStop}%
\end{thebibliography}%

\end{document}